\documentclass[final,5p,times,twocolumn]{elsarticle}

\usepackage{amsmath,amssymb,amsfonts,latexsym}
\usepackage{array,color}
\usepackage{tabularx}
\usepackage{multirow}
\usepackage{algorithm}
\usepackage{makecell}
\usepackage{CJK,graphicx}
\usepackage{booktabs}
\usepackage{subfigure}
\usepackage{txfonts}
\usepackage{slashbox}
\usepackage{color}
\usepackage{txfonts}
\usepackage{tabularx}
\usepackage{colortbl,multirow,hhline}
\usepackage{algorithmic}
\usepackage{epsfig}
\usepackage{float}
\usepackage{hyperref}
\usepackage{supertabular}

\journal{Journal of \LaTeX\ Templates}

\begin{document}

\begin{frontmatter}

\title{Quality Assurance Technologies of Big Data Applications:\\ A Systematic Literature Review}



\author[label1]{Pengcheng Zhang\corref{cor1}}
\address[label1]{College of Computer and Information,
Hohai University, Nanjing {\rm 211110}, P.R.China}

\cortext[cor1]{I am corresponding author}

\ead{pchzhang@hhu.edu.cn}
\ead[url]{author-one-homepage.com}

\author[label1]{Wennan Cao}

\author[label2]{Henry Muccini}
\address[label2]{Information Engineering and Computer Science and Mathematics Department, University of LAquila, Italy}
\ead{henry.muccini@univaq.it}

\begin{abstract}

Big data applications are currently used in many application domains, ranging from statistical applications to prediction systems and smart cities.
However, the quality of these applications is far from perfect, leading to a large amount of issues and problems. Consequently, assuring the overall quality for big data applications plays an increasingly important role. This paper aims at summarizing and assessing existing quality assurance (QA) technologies addressing quality issues in big data applications. We have conducted a systematic literature review (SLR) by searching major scientific databases, resulting in 83 primary and relevant studies on QA technologies for big data applications. The SLR results reveal the following main findings: 1) the impact of the big data attributes of \emph{volume}, \emph{velocity}, and \emph{variety} on the quality of big data applications; 2) the quality attributes that determine the quality for big data applications include \emph{correctness}, \emph{performance}, \emph{availability}, \emph{scalability}, \emph{reliability} and so on; 3) the existing QA technologies, including \emph{analysis}, \emph{specification}, \emph{model-driven architecture (MDA)}, \emph{verification}, \emph{fault tolerance},  \emph{testing}, \emph{monitoring} and \emph{fault \& failure prediction}; 4) existing strengths and limitations of each kind of QA technology;
5) the existing empirical evidence of each QA technology. This study provides a solid foundation for research on QA technologies of big data applications. However, many challenges of big data applications regarding quality still remain.

\end{abstract}


\end{frontmatter}

\section{Introduction}\label{sec:introduction}
The big data technology market grows at a 27\% compound annual growth rate (CAGR), and big data market opportunities will reach
 over $203$ billion \$ in 2020~\cite{2013Big,dai2019big}. Big data application systems~\cite{chen2014data,allam2019big}, abbreviated as big data applications, refer to the software systems that can collect, process, analyze or predict a large amount of data by means of different platforms, tools and mechanisms. Big data applications are now increasing being used in many areas, such as recommendation systems, monitoring systems, and statistical applications~\cite{Tao2016Quality,jan2019deep}. Big data applications are associated with the so-called 4V attributes, e.g., \emph{volume}, \emph{velocity}, \emph{variety} and \emph{veracity}~\cite{Hilbert2016Big}. Due to the large amount of generated data, the fast velocity of arriving data, and the various types of heterogeneous data, the quality of data is far from ideal, which makes the software quality of big data applications far from perfect~\cite{Laranjeiro2015A}. For example, due to the \emph{volume} and \emph{velocity} attributes~\cite{Anagnostopoulos2016Handling,Gudivada2015Big}, the generated data of big data applications are extremely large and increasing with high speed, which may affect \emph{data accuracy} and \emph{data timeliness}~\cite{Montagud2012A}, and consequently lead to software quality problems, such as \emph{performance} and \emph{availability} issues~\cite{Montagud2012A,nguyen2015impact}. Due to the huge \emph{variety} of heterogeneous data~\cite{Wang2016Towards,Bagriyanik2016Big}, data types and formats are increasingly rich, including \emph{structured}, \emph{semi-structured}, and \emph{unstructured}, which may affect \emph{data accessibility} and \emph{data scalability}, and hence lead to \emph{usability} and \emph{scalability} problems.

\newcommand{\tabincell}[2]{\begin{tabular}{@{}#1@{}}#2\end{tabular}}  
\renewcommand\arraystretch{1.8}
\begin{table*}[]
 \centering
\caption{\label{tab:test1}Key Findings and Implications of this Research}
 \begin{tabular}{p{0.8cm}p{7cm}p{8cm}}
  \toprule[1pt]
 \multicolumn{1}{l}{\textbf{Numbers}} &\multicolumn{1}{l}{\textbf{Key findings}} & \multicolumn{1}{l}{\textbf{Implications}}                                                                                                                                                         \\ \midrule[1pt]
F1  & {The three main big data attributes (\emph{volume} \emph{velocity}, and \emph{variety}) have a direct impact on the quality of the applications.}                                   & {The \emph{volume}, \emph{velocity} and \emph{variety} can affect data quality, and thus also affect the quality of big data applications.}                         \\ \hline
F2                    & {Some quality attributes, such as \emph{correctness}, \emph{performance}, \emph{availability}, \emph{scalability} and \emph{reliability}, can determine the quality of big data applications.}      & {Through our research, we can identify the most important quality attributes that state-of-the-art works address.}                            \\ \hline
F3                    & {Existing QA technologies include \emph{Analysis}, \emph{Specification}, \emph{Model-Driven Architecture (MDA)}, \emph{Testing}, \emph{Fault tolerance}, \emph{Fault and Failure Prediction}, \emph{Monitoring} and \emph{Verification}.} & {Surveying and summarizing existing quality assurance technologies for big data applications.} \\ \hline
F4                    & {Existing strengths and limitations of each kind of QA technique.}                                                                                             & {Through the systematic review, strengths and limitations of each kind of QA technique are discussed and compared.}                   \\ \hline
F5                    & {Existing empirical evidence of each kind of QA technique.}                                                                                             & {Validating the proposed QA technologies through real cases and providing a reference for big data practitioners.}                   \\ \bottomrule[1pt]
 \end{tabular}
\end{table*}

In general, quality assurance (QA) is a general way to detect or prevent mistakes or defects in manufactured software/products and avoid problems when solutions or services are delivered to customers~\cite{schulmeyer1992handbook}. However, compared with traditional software systems, big data applications raise new challenges for QA technologies due to the four big data attributes (for example, the \emph{velocity} of arriving data, and the \emph{volume} of data)~\cite{Gao2016Big}. Many scholars have illustrated current QA problems for big data applications~\cite{Lai2016Data}, [P39]. For example, it is a hard task to validate the \emph{performance}, \emph{availability} and \emph{accuracy} of a big data prediction system due to the large-scale data size and the feature of timeliness. Because of the \emph{volume} and \emph{variety} attributes, keeping big data recommendation systems scalable is very difficult. Therefore, QA technologies of big data applications is now becoming a hot research topic currently.

Compared with traditional applications, big data applications have the following special characteristics: a) statistical computation based on large-scale, diverse formats, with structured and non-structured data; b) machine learning and knowledge-based system evolution; c) intelligent decision-making with uncertainty and d) more complex visualization requirements. These new features of big data applications need novel QA technologies to ensure quality.
For example, compared with data in traditional applications (such as graphics, images, sounds, documents, etc.), there is a substantial amount of unstructured data in big data applications. These data are usually heterogeneous and lack of integration. Consequently, traditional testing processes lack testing methods for unstructured data and cannot adapt to the diversity of data processing requirements. Some novel QA technologies are urgently needed to solve these problems.

In the literature, many scholars have investigated the use of different QA technologies to assure the quality of big data applications~\cite{Zhou2015An,Juddoo2016Overview,Gao2016Big,Zhang2017A}.
Some papers have presented overviews on quality problems of big data applications. Zhou et al.~\cite{Zhou2015An} presented the first comprehensive study on the quality of the big data platform. For example, they have investigated the common symptoms, causes, and mitigation of quality issues, including hardware faults, code defects and so on. Juddoo~\cite{Juddoo2016Overview} et al. have systematically studied the challenges of data quality in the context of big data. Gao et al.~\cite{Gao2016Big} did a profound research on validation of big data and QA, including the basic concepts, issues, and validation process. They also discussed the big data QA focuses, challenges and requirements. Zhang et al.~\cite{Zhang2017A} introduced big data attributes and quality attributes, some quality assurance technologies like testing and monitoring were also discussed. Although these authors have proposed a few QA technologies for big data applications, publications on QA technologies for big data applications remain scattered in the literature, and this hampers the analysis of the advanced technologies and the identification of novel research directions. Therefore, a systematic study of QA technologies for big data applications is still necessary and critical.

In this paper, we provide an exhaustive survey of QA technologies that perform a significant role in big data applications, covering 83 papers published from Jan. 2012 to Dec. 2019.
The major purpose of this paper is to look into literature that is related to QA technologies for big data applications. Then, a comprehensive reference concerning challenges of QA technologies for big data applications is also proposed.
In summary, the major contributions of the paper are described in the following:
\begin{itemize}
  \item The elicitation of big data attributes, and the quality problems they introduce to big data applications;
  \item The identification of the most frequently used big data QA technologies, together with an analysis of their strengths and limitations;
  \item A discussion of existing strengths and limitations of each kind of QA technologies.
  \item The proposed QA technologies are generally validated through real cases, which provides a reference for big data practitioners.
\end{itemize}
Our research results in five overall main findings that are summarized in Table~\ref{tab:test1}.

The findings of this paper contribute general information for future research as the quality of big data applications becomes increasingly more important.
Existing QA technologies have a certain effect on the quality of big data applications; however, some challenges still exist, such as the lack of quantitative models and algorithms.

The rest of the paper is structured as follows. The next section reviews related background and previous studies. Section~\ref{sec_Research Method} describes our systematic approach for conducting the review. Section~\ref{sec_Results} reports the results of themes based on five research questions raised in Section~\ref{sec_Research Method}. Section~\ref{sec:DISCUSSION} provides the main findings of the survey and provides existing research challenges. Section~\ref{sec:THREATS TO VALIDITY} describes some threats in this study. Conclusions and future research directions are given in the final section.

\section{Related work}\label{sec_Related work}
To begin our study, we searched Google Scholar, Baidu Scholar, Bing Academic, IEEE, ACM and other search engines and databases (using the search strings: (systematic study OR literature review OR SLR OR SMS OR systematic literature review OR systematic mapping study) AND (Big data) AND (application OR system) AND (quality OR performance OR quality assurance OR QA))). We finally found that there is no systematic literature review (including a systematic mapping study, a systematic study, and a literature review) that focuses on QA technologies for big data applications. However, quality issues are prevalent in big data~\cite{informatics5020019}, and the quality of big data applications has attracted attention and been the focus of research in previous studies. In the following, we first try to describe all the relevant reviews that are truly related to the quality of big data applications.

Zhou et al.~\cite{Zhou2015An} firstly present comprehensive study on the quality of the big data platform. They have investigated the familiar symptoms, causes, and mitigation of quality problems. In addition, big data computing also presents different types of problems, including hardware failure, code defects and so on. Their discovery is of great significance to the design and maintenance of big data platforms in the future.

Juddoo et al.~\cite{Juddoo2016Overview} systematically study the challenges of data quality in the context of big data. They mainly analyze and propose the data quality technologies that would be more suitable for big data in a general context. Their goal is to probe diverse components and activities forming part of data quality management, metrics, dimensions, data quality rules, data profiling, and data cleansing. In addition, the volume, velocity, and variety of data may make it impossible to determine the data quality rules. They believe that the measurement of big data attributes is very important to the users' decision-making. Finally, they also list existing data quality challenges.

Gao and Tao~\cite{Tao2016Quality,Gao2016Big} first provide detailed discussions for QA problems and big data validation, including the basic concepts and key points. Then they discuss big data applications influenced by big data features. Furthermore, they also discuss big data validation processes, including data collection, data cleaning, data cleansing, data analysis, etc. In addition, they summarize the big data QA issues, challenges and needs.

Zhang et al.~\cite{Zhang2017A} further consider QA of big data applications, combined QA technologies with big data attributes, and explore the big data 4V attributes of existing QA technologies.

Liu et al.~\cite{Liu2016Rethinking} point out and summarize the issues faced by big data research in data collection, processing and analysis in the current big data area, including uncertain data collection, incomplete information, and big data noise, representability, consistency, reliability and so on.

To sum up, big data applications offer many opportunities to adjust businesses and enhance promotion models.
In addition, big data applications can also help governments perform accurate prediction, such as weather forecasts, preventing natural disasters, and developing appropriate policies to improve the quality of human life. The survey of existing literature (such as paper~\cite{Lai2016Data, Zhou2015An, Juddoo2016Overview, Zhang2017A, Liu2016Rethinking}) shows that there have been many studies to introduce big data QA, but little scientific research has focused on comprehending, defining, classifying and communicating QA technologies for big data applications. Consequently, there is no definite way to address QA of big data applications. As a result, it is necessary to conduct a systematic study of QA technologies for big data applications.

\section{Research Method}\label{sec_Research Method}
\begin{figure*}[t]
  \centering
  \includegraphics[width=0.8\linewidth]{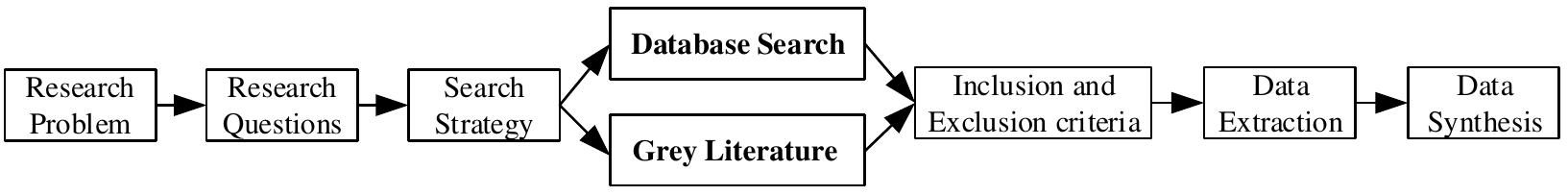}\\
  \caption{SLR Protocol}\label{fig1}
\end{figure*}

In this work, a systematic literature review (SLR) approach proposed by Kitchenham et al.~\cite{Kitchenham2009Systematic} is used to extract QA technologies for big data applications and related questions. Based on the SLR and our research problem, research steps can be raised as shown in Fig.~\ref{fig1}. Through these research steps, we can obtain the desired results.

\subsection{Research Questions}\label{sec:Research Questions}
We used the Goal-Question-Metric (GQM) perspectives (i.e., purpose, issue, object, and viewpoint)\cite{Kitchenham2004Procedures} to draw up the aim of this study. The result of the application of the Goal-Question-Metric approach is the specification of a measurement system targeting the given set of problems and a set of rules for understanding the measurement data \cite{Basili1994The}. Table~\ref{tab:test2} provides the purpose, issue, object and viewpoint of the research topics.

Research questions can usually help us to perform an in-depth study and achieve purposeful research. Based on this research, there are five research questions. Table~\ref{tab:test3} shows the research questions that we translated from Table~\ref{tab:test2}. The five research questions are in Table~\ref{tab:test3}, and their subresearch questions are detailed below. For each research question, we also propose the primary objective of the investigation.

\noindent \textbf{RQ1:} How do the big data attributes affect the quality of big data applications? Generally, this problem involves all big data attributes.

\noindent \textbf{Objective:} Discuss the influence of the big data attributes on the quality of big data applications.

\begin{table}[htbp]
 \centering
\caption{\label{tab:test2}Goal of this Research}
\begin{tabular}{p{0.4cm}p{1.2cm}p{6.2cm}}
  \toprule[1pt]
  Goal & \tabincell{c}{Purpose \\Issue \\Object \\viewpoint}
 & \tabincell{l}{Identify, analyze and extract QA technologies \\for big data applications, and then understand \\the features and challenges of technologies \\in existence from a researcher's viewpoint.} \\
  \midrule[1pt]
\end{tabular}
\end{table}

\begin{table}[h]
 \centering
\caption{\label{tab:test3}Research Questions}
\begin{tabular}{p{0.6cm}p{7.2cm}}
  \toprule[1pt]
  \textbf{ID} & \textbf{Research Question} \\\midrule[1pt]
  RQ1 & {How do the big data attributes affect the quality of big data applications?} \\\hline
  RQ2 & {Which kind of important quality attributes do big data applications need to guarantee?} \\\hline
  RQ3 & {Which kinds of technologies are used to guarantee the quality of big data applications?} \\\hline
  RQ4 & {What are the strengths and limitations of the proposed technologies?} \\\hline
  RQ5 & {What are the real cases of using the proposed technologies?} \\\bottomrule[1pt]
\end{tabular}
\end{table}

\noindent \textbf{RQ2:} Which kind of important quality attributes do big data applications need to ensure?

\noindent \textbf{Objective:} Identify and classify existing common quality attributes and understand the implication of them.

\noindent \textbf{RQ3:} Which kinds of technologies are used to guarantee the quality of big data applications?

\noindent \textbf{Objective:} Identify and classify existing QA technologies and understand the effect of them.

\noindent \textbf{RQ4:} What are the strengths and limitations of the proposed technologies?

\noindent \textbf{Objective:} Analyze the strengths and limitations of those QA technologies.

\noindent \textbf{RQ5:} What are the real cases of using the proposed technologies?

\noindent \textbf{Objective:} Validate the proposed QA technologies through real cases and provide a reference for big data practitioners.

\subsection{Search Strategy}\label{sec:Search Strategy}
The goal of this systematic review is thoroughly examining the literature on QA technologies for big data applications. Three main phases of SLR are presented by EBSE (evidence-based software engineering)~\cite{Kitchenham2004Evidence} guidelines that include planning, execution, and reporting results. Moreover, the search strategy is an indispensable part and consists of two different stages.

\noindent \textbf{Stage 1: Database search.}

Before we carried out automatic searches, the first step was the definition and validation of the search string to be used for automated search. This process started with pilot searches on seven databases as shown in Table~\ref{tab:test4}. We combined different keywords that are related to research questions. Table~\ref{tab:test5} shows the search terms we used in the seven databases, and the search string is defined in the following:

\begin{table}[]
 \centering
\caption{\label{tab:test4}Studies Resource}
\begin{tabular}{p{3cm}p{5cm}}
  \toprule[1pt]
  \textbf{Source} & \textbf{Address} \\\midrule[1pt]
  ACM Digital Library & {http://dl.acm.org/} \\\hline
  IEEE Xplore\\
  Digital Library & {http://ieeexplore.ieee.org/} \\\hline
  Springer Link & {http://link.springer.com/} \\\hline
  Science Direct & {http://www.sciencedirect.com/} \\\hline
  Scopus & {http://www.scopus.com/} \\\hline
  Engineering Village & http://www.engineeringvillage.com/ \\\hline
  ISI Web of Science & http://isiknowledge.com/ \\
  \bottomrule[1pt]
\end{tabular}
\end{table}

\begin{table}[]
 \centering
\caption{\label{tab:test5}Search Terms}
\begin{tabular}{cl}
  \toprule[1pt]
  \textbf{Search ID} & \textbf{Database Search Terms} \\\midrule[1pt]
  {a} & {big data}\\\hline
  {b} & {application}\\\hline
  {c} & {system} \\\hline
  {d} & {quality} \\\hline
  {e} & {performance}\\\hline
  {f} & {quality assurance}\\\hline
  {g} & {QA}\\
  \bottomrule[1pt]
\end{tabular}
\end{table}

\noindent (a AND (b OR c) AND (d OR e OR f OR g)) IN (Title or Abstract or Keyword).

We used a ``quasi-gold standard"\cite{Zhang2010On} to validate and guarantee the search string,. We use IEEE and ACM libraries as representative search engines to perform automatic search and refine the search string until all the search items meet the requirements and the number of remaining papers was minimal. Then, we use the defined search string to carry out automatic searches. We choose ACM Digital Library, IEEE Xplore Digital Library, Engineering Village, Springer Link, Scopus, ISI Web of Science and Science Direct because these seven databases are the largest and most complete scientific databases that include computer science. We manually downloaded and searched the proceedings if venues not included in the digital libraries. After the automatic search, a total of 3328 papers were collected.

\noindent \textbf{Stage 2: Grey literature.}

To cover grey literature, some alternative sources were investigated as follows:

\begin{itemize}
  \item Google Scholar

  In order to adapt the search terms to Google Scholar and improve the efficiency of the search process, search terms were slightly modified. We searched and collected 1220 papers according to the following search terms:
  \\-- (big data AND (application OR system) AND ((quality OR performance OR QA) OR testing OR analysis OR verification OR validation))
  \\-- (big data AND (application OR system) AND (quality OR performance OR QA) AND (technique OR method))
  \\-- (big data AND (application OR system) AND (quality OR performance OR QA) AND (problem OR issue OR question))
  \item Checking the personal websites of all the authors of primary studies, in search of other related sources (e.g., unpublished or latest progress).
\end{itemize}

Through two stages, we found 4548 related papers. Only 102 articles met the selection strategy (discussed below) and are chosen in the next stage. Then, we scanned all the related results according to the snowball method~\cite{Goodman1961Snowball}, and we referred to the references cited by the selected paper and include them if they are appropriate. We expanded the number of papers to 121; for example, we used this technique to find P72, which corresponds to our research questions from the references in P10.

To better manage the paper data, we used NoteExpress~\footnote{https://noteexpress.apponic.com/}, which is a professional-level document retrieval and management system. Its core functions cover all aspects of ``knowledge acquisition, management, application, and mining". It is a perfect tool for academic research and knowledge management. However, the number of these results is too large and therefore detrimental to the study.
Consequently, we filtered the results by using the selection strategy described in the next section.

\subsection{Selection Strategy}\label{sec:Selection Strategy}
In this subsection, we focus on the selection of research literature. According to the search strategy, much of the returned literature is unnecessary. It is essential to define the selection criteria (inclusion and exclusion criteria) for selecting the related literature. We describe each step of our selection process in the following:

\noindent \textbf{Step 1:} Combination and duplicates removal. In this step, we sort out the results that we obtain from stage 1 and stage 2 and remove the duplicate content.

\noindent \textbf{Step 2:} Selection of studies. In this step, the main objective is to filter all the selected literature in the light of a set of rigorous inclusion and exclusion criteria. There are five inclusion and four exclusion selection criteria we have defined as described below.

\noindent \textbf{Step 3:} Exclusion of literature during data extraction. When we read a study carefully, it can be selected or rejected according to the inclusion and exclusion criteria in the end.

When all inclusion criteria are met, the study is selected; otherwise, it is discarded if any exclusion criteria are met. According to the research questions and research purposes, we identified the following inclusion and exclusion criteria.

\noindent \textbf{Inclusion criteria:} a study should be chosen if it satisfies each inclusion criteria.

 \noindent 1) The study of the literature focuses on the quality of big data applications or big data systems, in order to be aligned with the theme of our study.

 \noindent 2) One or more of our research questions must be directly answered.

 \noindent 3) The selected literature must be in English.

 \noindent 4) The literature must consist of journal papers or papers published as part of conference or workshop proceedings.

\begin{figure}[ht]
  \centering
  \includegraphics[width=1\linewidth]{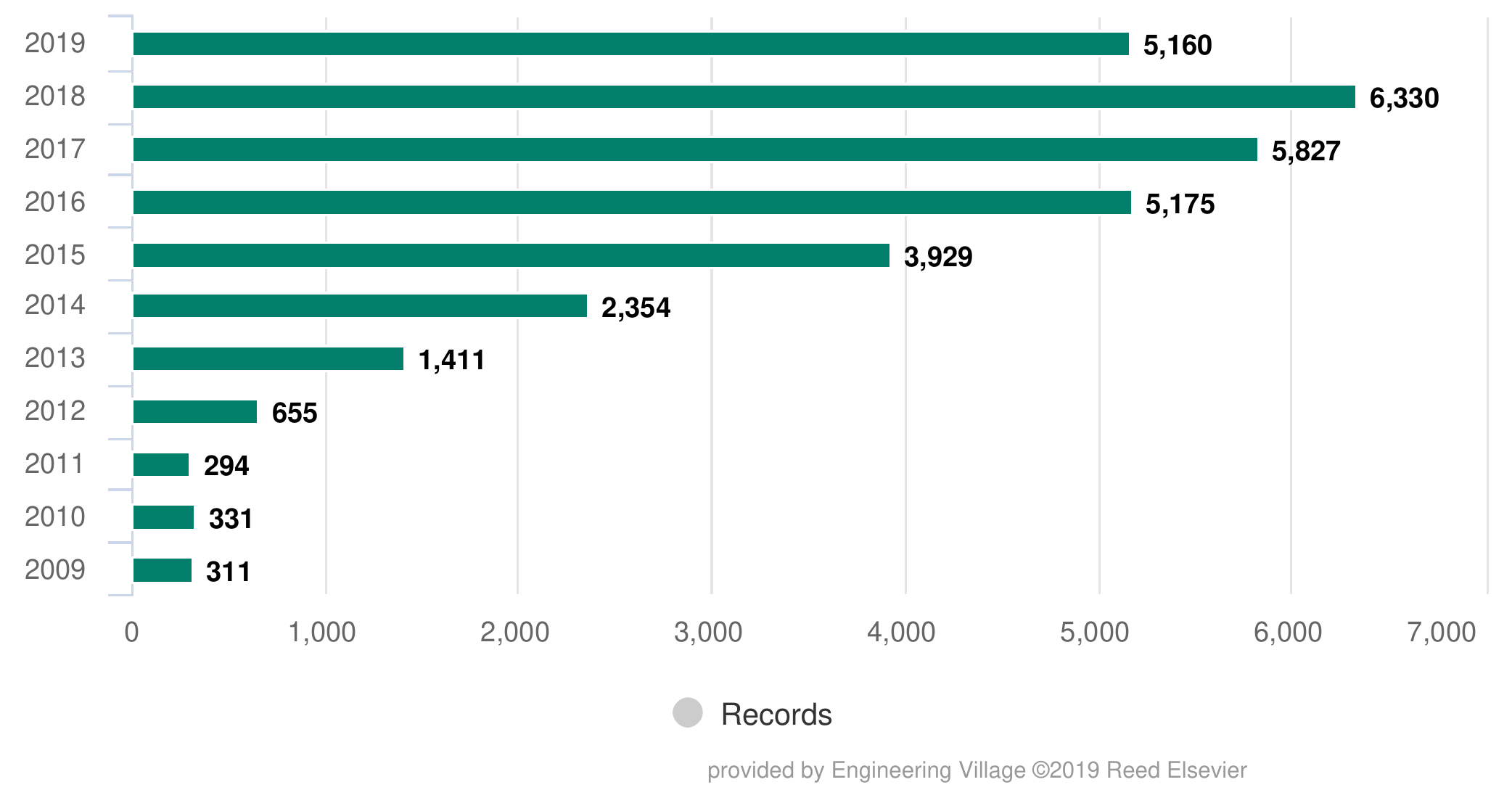}\\
  \caption{Distribution of Papers in EI until Dec 2019}\label{fig2}
\end{figure}

 \noindent 5) Studies are published in or after 2012. From a simple search (for which we use the search item "big data application") in the EI (engineering index) search library, we can see that most of the papers on big data applications or big data systems are published after 2011, as shown in Fig.~\ref{fig2}. The literature published before 2012 rarely takes into account the quality of big data applications or systems. By reading the relevant literature abstracts, we found that these earlier papers were not relevant to our subject, so we excluded them.

The main objective of our study is to determine the current technologies of ensuring the quality of big data applications and the challenges associated with the quality of big data applications. This means that the content of the article should be related to the research questions of this paper.

\noindent \textbf{Exclusion criteria:} a study should be discarded if it satisfies any one of the following exclusion criteria.

\noindent 1) It is related to big data but not related to the quality of big data applications. Our goal is to study the quality of big data applications or services, rather than the data quality of big data, although data quality can affect application quality.

\noindent 2) It does not explicitly discuss the quality of big data applications and the impact of big data applications or quality factors of big data systems.

\noindent 3) Duplicated literature. Many articles have been included in different databases, and the search results contain repeated articles. For conference papers that meet our selection criteria but are also extended to journal publications, we choose journal publications because they are more comprehensive in content.

\noindent 4) Studies that are not related to the research questions.

 \begin{figure}
  \centering
  \includegraphics[width=1\linewidth]{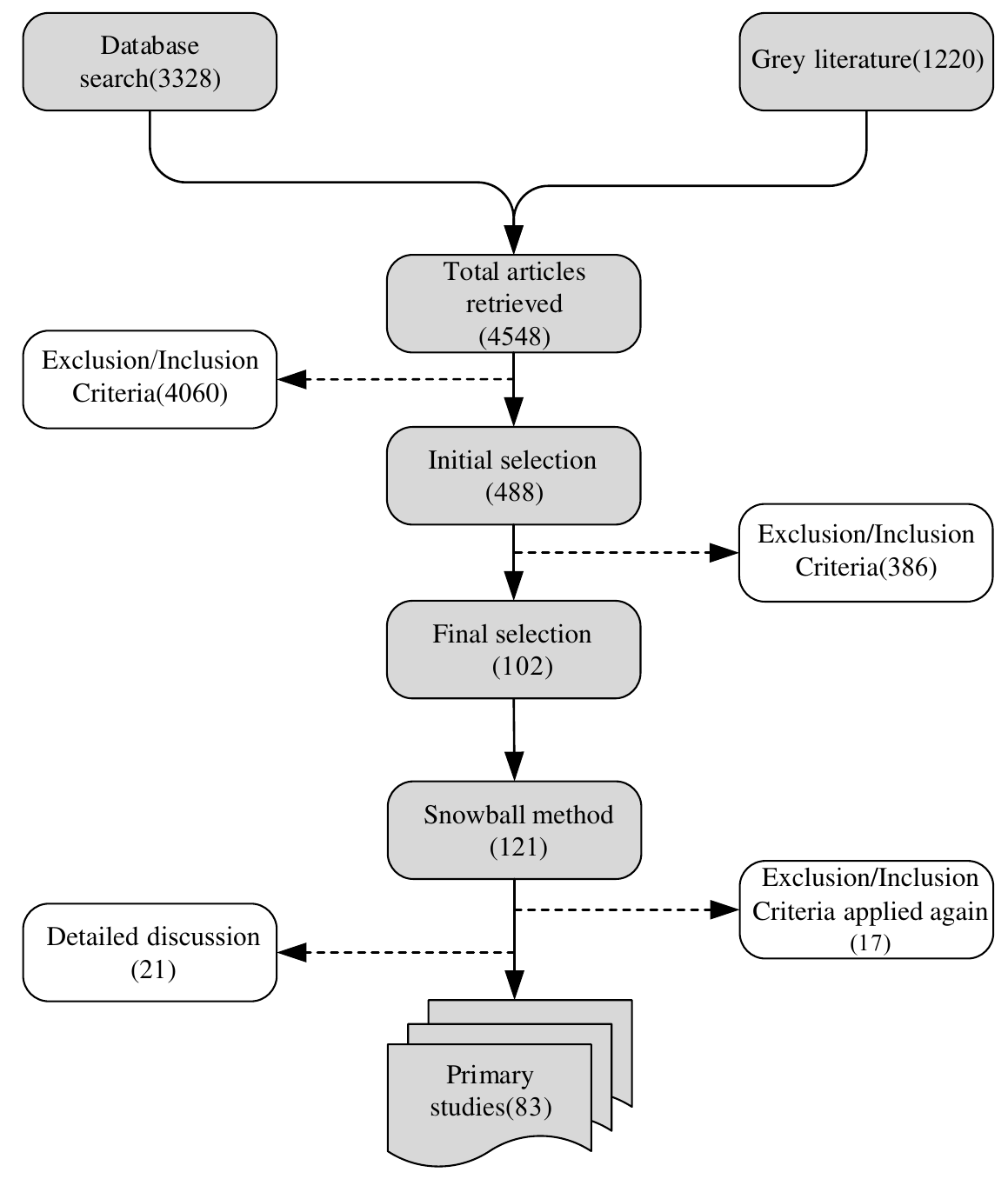}\\
  \caption{The Search Process}\label{fig3}
\end{figure}

Inclusion criteria and exclusion criteria are complementary. Consequently, both the inclusion and exclusion criteria are considered. In this way, we can achieve the desired results. A detailed process of identifying relevant literature is presented in Figure~\ref{fig3}. Obviously, the analysis of all the literature presents a certain degree of difficulty. First, by applying these inclusion and exclusion criteria, two researchers separately read the abstracts of all studies selected in the previous step to avoid prejudice as much as possible. Of the initial studies, 488 were selected in this process. Second, for the final selection, we read the entire initial papers and then selected 102 studies. Third, we expanded the number of final studies to 121 according to our snowball method. Conflicts were resolved by extensive discussion. We excluded numbers of papers because they were not related and had 83 primary studies at the end of this step. Details of the selected papers are shown in Table 17 of Appendix B.

In the process, the first author and the second author worked together to develop research questions and search strategies, and the second author and four students of the first author executed the search plan together. During the process of finalizing the primary articles, all members of the group had a detailed discussion on whether the articles excluded by only few researchers are in line with our research topic.

\subsection{Quality Assessment}\label{sec:Quality Assessment}

After screening the final primary studies by inclusion and exclusion criteria, the criteria for the quality of the study were determined according to the guidelines proposed by Kitchenham and Charters\cite{Kitchenham07guidelinesfor}. The corresponding quality checklist is shown in Table~\ref{tab:14}. The table includes 12 questions that consider four research quality dimensions, including research design, behavior, analysis, and conclusions. For each quality item we set a value of 1 if the authors put forward an explicit description, 0.5 if there is a vague description, and 0 if there is no description at all.
The author and his research assistant applied the quality assessment method to each major article, compared the results, and discussed any differences until a consensus was reached. We scored each possible answer for each question in the main article and converted it into a percentage after coming to an agreement.
The presentation quality assessment results of the preliminary study indicate that most studies have a deep description of the problem and its background, and most studies have fully and clearly described the contributions and insights. Nevertheless, some studies do not describe the specific division of labor in the method introduction, and there is a lack of discussion of the limitations of the proposed method. However, the total average score of 8.8 out of 12 indicates that the quality of the research report is good, supporting the validity of the extracted data and the conclusions drawn therefrom.

\renewcommand\arraystretch{1}
\begin{table*}[t]
 \centering
 \caption{\label{tab:14}Quality Assessment Questions and Results}
\begin{tabular}{lllll}
  \toprule[1pt]
\multirow{2}{*}{ID} & \multicolumn{1}{l}{\multirow{2}{*}{Question}}  & \multicolumn{3}{c}{Percentage}  \\ \cline{3-5} & \multicolumn{1}{l}{}                                                                 & \multicolumn{1}{l}{Yes} & \multicolumn{1}{l}{partially} & \multicolumn{1}{l}{No} \\ \midrule[1pt]
\multicolumn{5}{l}{\textbf{Design}}                                                                                                           \\\midrule[1pt]
Q1                  & Are the aims of the study clearly stated?                                             &100\%\        &0\%\           &0\%\           \\
Q2                  & Are the chosen quality attributes distinctly stated and defined?                     &55.4\%\       &43.2\%\        &0.4\%\         \\
Q3                  & Was the sample size reasonable?                                                       &32.4\%\       &45.9\%\        &21.7\%\        \\\midrule[1pt]
\textbf{Conduct}    &                                                                                       &              &               &               \\\midrule[1pt]
Q4                  & Are research methods adequately described?                                            &90.5\%\       &9.5\%\         &0\%\           \\
Q5                  & Are the datasets completely described (source, size, and programming languages)?      &32.4\%\       &43.2\%\        &24.4\%\        \\
Q6                  & Are the observation units or research participants described in the study?            &2.7\%\        &0\%\           &97.3\%\        \\\midrule[1pt]
\textbf{Analysis}   &                                                                                       &              &               &               \\\midrule[1pt]
Q7                  & Is the purpose of the analysis clear?                                                 &98.6\%\       &1.4\%\         &0\%\           \\
Q8                  & Are the statistical methods described?                                                &14.9\%\       &5.4\%\         &79.7\%\        \\
Q9                  & Is the statistical significance of the results reported?                              &14.9\%\       &5.4\%\         &79.7\%\        \\\midrule[1pt]
\textbf{Conclusion} &                                                                                       &              &               &               \\\midrule[1pt]
Q10                 & Are the results compared with other methods?                                          &27.0\%\       &1.4\%\         &71.6\%\        \\
Q11                 & Do the results support the conclusions?                                               &100\%\        &0\%\           &0\%\           \\
Q12                 & Are validity threats discussed?                                                       &18.9\%\       &28.4\%\        &52.7\%\        \\\bottomrule[1pt]
\end{tabular}
\end{table*}

\subsection{Data Extraction}\label{sec:Data Extraction}
The goal of this step is to design forms to identify and collect useful and relevant information from the selected primary studies so that it can answer our research questions proposed in Section~\ref{sec:Research Questions}. To carry out an in-depth analysis, we can apply the data extraction form to all selected primary studies. Table~\ref{tab:test6} shows the data extraction form. According to the data extraction form, we collect specific information in an excel file \footnote{https://github.com/QXL4515/QA-techniques-for-big-data-application}. In this process, the first author and the second author jointly developed an information extraction strategy to lay the foundation for subsequent analysis. In addition, the third author validated and confirmed this research strategy.

\subsection{Data Synthesis}\label{sec:Data Synthesis}
Data synthesis is used to collect and summarize the data extracted from primary studies. Moreover, the main goal is to understand, analyze and extract current QA technologies for big data applications. Our data synthesis is specifically divided into two main phases.

\textbf{Phase 1:} We analyze the extracted data (most of which are included in Table~\ref{tab:test6} and some are indispensable in the research process) to determine the trends and collect information about our research questions and record them. In addition, we classify and analyze articles according to the research questions proposed in Section~\ref{sec:Research Questions}.

\textbf{Phase 2:} We classify the literature according to different research questions. The most important task is to classify the articles according to the QA technologies through the relevant analysis.

\renewcommand\arraystretch{1.5}
\begin{table*}[]
 \centering
\caption{\label{tab:test6}Data Extraction Form}
\begin{tabular}{cllc}
  \toprule[1pt]
  \textbf{Data Item} & \textbf{Extracted data} & \textbf{Description} & \textbf{Type} \\\midrule[1pt]
  1 & \tabincell{c}{Study title} & \tabincell{c}{Reflect the relevant research direction} & \tabincell{c}{Whole research} \\\hline
  2 & \tabincell{c}{Publication year} & \tabincell{c}{Indicating the trend of research} & \tabincell{c}{Whole research} \\\hline
  3 & \tabincell{c}{Journal/Conference} & \tabincell{c}{The type of study: the conference or the journal} & \tabincell{c}{Whole research} \\\hline
  4 & \tabincell{c}{Authors} & \tabincell{c}{Research author's other relevant studies} & \tabincell{c}{Whole research} \\\hline
  5 & \tabincell{c}{Context study} & \tabincell{c}{Understand the full text} & \tabincell{c}{Whole research} \\\hline
  6 & \tabincell{c}{Existing challenges} & \tabincell{l}{The limitations of the approaches and the challenges \\of big data application} & \tabincell{c}{Whole research} \\\hline
  7 & \tabincell{c}{Big data attributes} & \tabincell{c}{Related 4V attributes} & \tabincell{c}{RQ1} \\\hline
  8 & \tabincell{c}{Quality requirements} & \tabincell{c}{The attributes of the demand} & \tabincell{c}{RQ2} \\\hline
  9 & \tabincell{c}{Quality attributes} & \tabincell{c}{Quality attributes of big data application} & \tabincell{c}{RQ2} \\\hline
  10 & \tabincell{c}{Technology} & \tabincell{c}{Application technology} & \tabincell{c}{RQ3} \\\hline
  11 & \tabincell{c}{Quality assurance technologies} & \tabincell{c}{Application domain} & \tabincell{c}{RQ3} \\\hline
  12 & \tabincell{c}{Experimental results} & \tabincell{c}{The effectiveness of the methods} & \tabincell{c}{RQ3} \\\hline
  13 & \tabincell{c}{Strengths} & \tabincell{c}{The advantages of the approaches} & \tabincell{c}{RQ4} \\\hline
  14 & \tabincell{c}{Empirical Evidence} & \tabincell{c}{Real cases of the methods} & \tabincell{c}{RQ5} \\\bottomrule[1pt]
\end{tabular}
\end{table*}

\section{Results}\label{sec_Results}
This section, by deeply analyzing the primary studies listed in Appendix B, provides an answer to the five research questions presented in Section~\ref{sec_Research Method}. For readability, we use [Px] refers to a surveyed paper listed in Appendix B.

In addition, Figures~\ref{fig4}, ~\ref{fig5}, and~\ref{fig6} provide some simple statistics. Figure~\ref{fig4} presents how our primary studies are distributed over the years. Figure~\ref{fig5} groups the primary studies according to the type of publication. Figure~\ref{fig6} counts the number or studies retrieved from different databases.

While Section~\ref{sec:Results Overview} provides an overview of the main concepts discussed in this section, Sections~\ref{sec:Identify the Effect of Big Data Properties Using SLR (RQ1)} to \ref{sec:Empirical Evidence (RQ5)} report the answer to the research questions.

\begin{figure}
  \centering
  \includegraphics[width=1\linewidth]{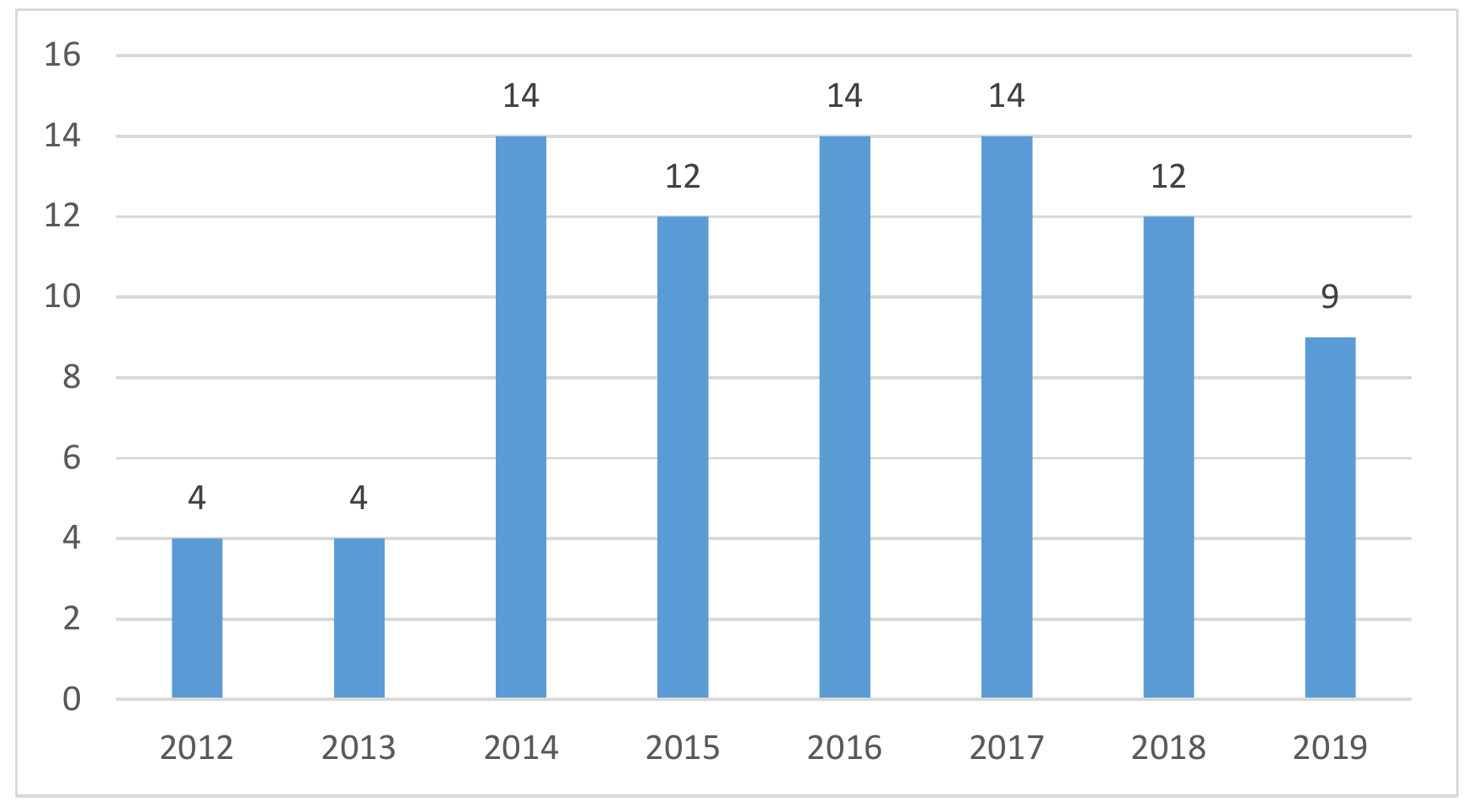}\\
  \caption{Distribution of Papers during Years 2012-2019}\label{fig4}
\end{figure}

\begin{figure}[]
  \centering
  \includegraphics[width=1\linewidth]{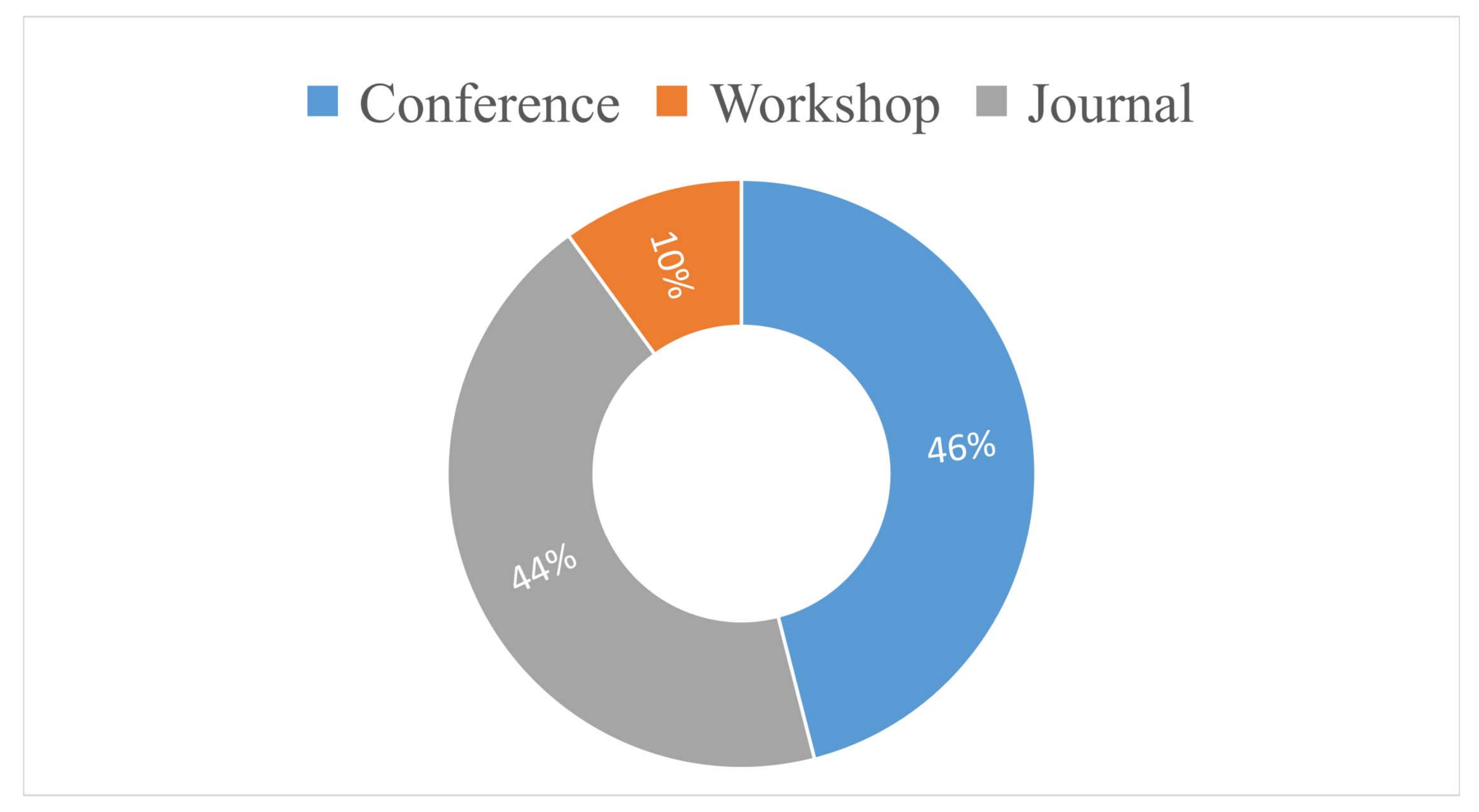}\\
  \caption{Distribution of the Types of Literature }\label{fig5}
\end{figure}

\begin{figure}[]
  \centering
  \includegraphics[width=1\linewidth]{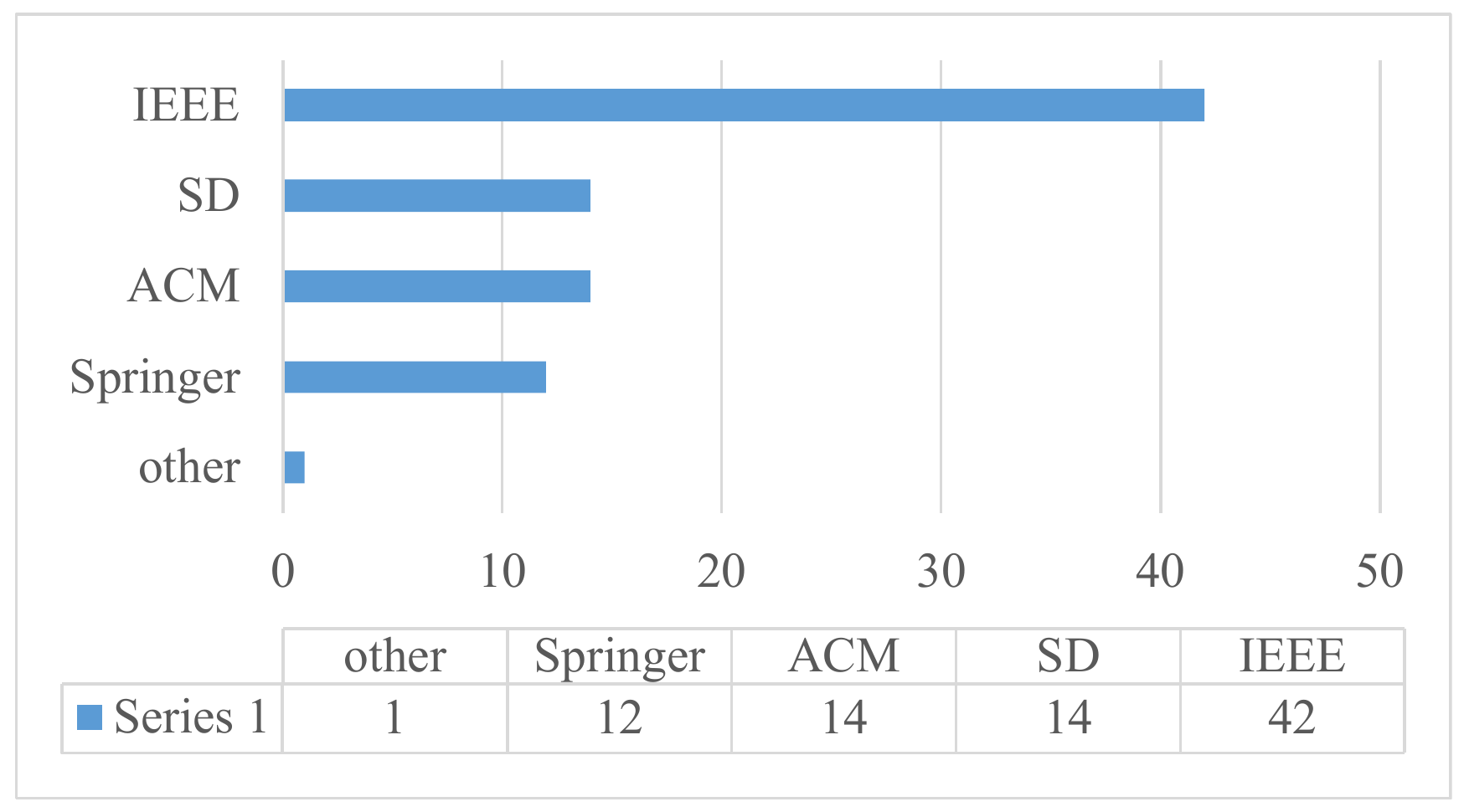}\\
  \caption{The Number Papers from Different Databases}\label{fig6}
\end{figure}

\subsection{Overview of the main concepts}\label{sec:Results Overview}

While answering the five research questions identified in previous sections, we will utilize and correlate three different dimensions: {\em big data attributes}, {\em data quality parameters}, and {\em software quality attributes}. (The output of this analysis is reported in Section \ref{sec:Existing Challenges (RQ5)} and Figure \ref{fig9}).

{\em Big data attributes:} big data applications are associated with the so-called 4V attributes, e.g., \emph{volume}, \emph{velocity}, \emph{variety} and \emph{veracity}~\cite{Hilbert2016Big}. In this study, we take into account only three of the 4V big data attributes (excluding the \emph{veracity} one) for the following reasons: First, through the initial reading of the literature, many papers are not concerned about the \emph{veracity}. Second, big data currently have multi-V attributes; only three attributes (\emph{volume}, \emph{variety} and \emph{velocity}) are recognized extensively\cite{Aggarwal2016Identification}\cite{Fasel2014Potentials}.

{\em Data quality parameters}: data quality parameters describe the measure of the quality of data. Since data are an increasingly vital part of applications, data quality becomes an important concern. Poor data quality could affect enterprise revenue, waste company resources, introduce lost productivity, and even lead to wrong business decisions~\cite{Gao2016Big}. According to the Experian Data Quality global benchmark report\footnote{Erin Haselkorn, New Experian Data Quality research shows inaccurate data preventing desired customer insight", Posted on Jan 29 2015 at URL http://www.experian.com/blogs/news/2015/01/29/data-quality-research-study/}, U.S. organizations claim that 32 percent of their data is wrong on average. Since data quality parameters are not universally agreed upon, we extracted them by analyzing papers~\cite{Gao2016Big,becker2015big,Clarke2016Big}.

{\em Software quality attributes}: software quality attributes describe the attributes that software systems shall expose. We start from the list provided in the ISO/IEC 25010:2011 standard and select those quality attributes that are mostly recurrent in the primary studies. Improved software quality attributes definitions are as follows:

 \noindent 1) A quality model consists of five characteristics (correctness, performance, availability, scalability, and reliability) that relate to the outcome of the interaction when a product is used in a particular context of use. This system model is applicable to the complete human-computer system, including both computer systems in use and software products in use.

 \noindent 2) A product quality model composed of eight characteristics (specification, analysis, MDA, fault tolerance, verification, testing, monitoring, fault and failure prediction) that relate to static attributes of software and dynamic attributes of the computer system. The model is applicable to both computer systems and software products.

\subsection{Identify the Effect of Big Data Properties (RQ1)}\label{sec:Identify the Effect of Big Data Properties Using SLR (RQ1)}

The goal of this section is to answer RQ1 ({\em how do the big data attributes affect the quality of big data applications?}). Table~\ref{tab:test7} lists the general challenges proposed by big data attributes. These challenges produce great difficulties regarding the quality of the data, thus affecting the quality of big data applications.

The \emph{volume} data property poses storage and scale challenges. The size of data sets used in industrial environments is huge, usually measured in terabytes, or even exabytes. Various types of data sets need a huge space to be stored and processed [P40]. Application \emph{performance} will decline as data \emph{volume} grows. When the amount of data reaches a certain size, the application crashes and cannot provide mission services [P41]. Therefore, massive amounts of data will inevitably affect the processing \emph{performance} of big data applications.

The \emph{velocity} data property poses fast analysis and processing challenges. With the flood of data generated quickly from smart phones and sensors, the new trend of big data analysis has shifted the focus to ``what can we do with data"~\cite{article}. Rapid analysis and processing of data need to be considered [P52]. The data generates and processes quickly and is therefore prone to errors. Mapping MapReduce frameworks to the cloud architecture became imperative in recent years because of the need to manage large data sets in a fast, reliable (and as cheap as possible) way [P22].

The \emph{variety} property poses the heterogeneity challenge, which leads to higher requirements on the data processing capacity of big data applications~\cite{Lai2016Data,Zhou2015An}. The increasing amount of sensors that are deployed on the Internet makes the generated data complex. It is impossible for human beings to write every rule for each type of data to identify relevant information. As a result, most of the events in these data are unknown, abnormal and indescribable. The collection, analysis, auditing, management and testing of such a complex amount of data by industry, researchers, government and media has become a major problem [P39].

\begin{table*}[!htbp]
 \centering
\caption{\label{tab:test7}The Challenge and Description of Big Data Properties}

\begin{tabular}{p{1.2cm}p{1.8cm}p{13.8cm}}
  \toprule[1pt]
  \textbf{Properties} & \textbf{Challenge} & \textbf{Description}\\\midrule[1pt]
  Volume & \tabincell{l}{Storage/Scale} & \tabincell{c}{The data scale has a significant effect on the performance of big data applications.}\\\hline
  Velocity & \tabincell{l}{Fast Processing} & \tabincell{c}{The data generates quickly, the data are processed quickly, and the data are prone to errors.}\\\hline
  Variety & \tabincell{l}{Heterogeneity} & \tabincell{c}{Multitype data need higher requirements on the data processing capacity of big data applications.}\\
  \bottomrule[1pt]
\end{tabular}
\end{table*}

\begin{table}[]
 \centering
\caption{\label{tab:test8}Distribution of the Relation between Big Data Properties and Data Quality Parameters}
\begin{tabular}{lll}
  \toprule[1pt]
  \textbf{Relation} & \textbf{Paper ID}\\\midrule[1pt]
  Volume-Data Correctness & \tabincell{c}{[P39], [P42]} \\\hline
  Volume-Data Completeness& \tabincell{c}{[P23], [P54]} \\\hline
  Volume-Data Timeliness & \tabincell{c}{[P41], [P53], [P24]} \\\hline
  Variety-Data Accuracy& \tabincell{c}{[P42], [P55]} \\\hline
  Variety-Data Consistency& \tabincell{c}{[P39], [P55], [P42], [P33]} \\\hline
  Velocity-Data Timeliness& \tabincell{c}{[P39], [P65], [P23], [P56]} \\\hline
  Velocity-Data Correctness& \tabincell{c}{[P54], [P7]} \\
  \bottomrule[1pt]
\end{tabular}
\end{table}

To answer RQ1, we extracted the relationships existing between the big data attributes and the data quality parameters. Table~\ref{tab:test8} identifies the primary studies discussing the relationship between couples of big data attributes and data quality parameters. The relationships that we found on the primary studies are reported in Table~\ref{tab:test8} and discussed below.

\begin{itemize}
  \item \textbf{Volume-Data Correctness:} According to [P39] and [P42], the larger the \emph{volume} of the data, the greater the probability that the data will be modified, deleted, and so on. In other words, a large volume of data has a high probability of errors in transmission, processing, and storage.

  \item \textbf{Volume-Data Completeness:} Data completeness is a quantitative measurement that is used to evaluate how much valid analytical data are obtained compared to the planned number~\cite{Gao2016Big}. Data completeness is usually expressed as a percentage of usable analytical data. In general, increased data reduces data completeness [P23][P54].

  \item \textbf{Volume-Data Timeliness:} In the era of big data, people are not only concerned with the size of data but also with the way to process data. Because the amount of data to be processed is too large, the business needs and competitive pressures require a real-time and effective data processing [P24], and response time is critical in most situations [P53]. Thus how to deal with a massive amount of data in a very short time is a vast challenge. If these data cannot be processed in a timely manner, the value of these data will decrease and the original goal of building big data systems will be lost [P41].

  \item \textbf{Variety-Data Accuracy:} For big data applications, the sources of data are varied, including \emph{structured}, \emph{semi-structured}, and \emph{unstructured} data. A part of these data has no statistical significance, which greatly influences the accuracy of big data application results [P55]. In addition, the contents of the database became corrupted by erroneous programs storing incorrect values and deleting essential records. It is hard to recognize such quality erosion in large databases, but over time, it spreads similar to a cancerous infection, causing ever-increasing big data system failures. Thus, not only data quality but also the quality of applications suffers under erosion [P42].

  \item \textbf{Variety-Data Consistency:} Data consistency is useful to evaluate the consistency of given data sets from different perspectives [P33]. The consistency of various data has a positive impact on the content validity and consistency of large data application systems [P42], [P55]. \emph{Unstructured} data will produce the consistency problem such as continuous availability and data security issues in big data applications [P39].

  \item \textbf{Velocity-Data Timeliness:} How to effectively handle high-speed data transmission and ensure the timeliness of data processing are very important. These data must be analyzed in time because the \emph{velocity} of data generation is very quick,[P65], [P23]. Generally, the greater the \emph{velocity} with which data can be analyzed is, the larger the profit for the organization [P39]. Low-speed data may result in the fact that the big data systems are unable to respond effectively to any negative change (speed)[P56]. Therefore, the \emph{velocity} engenders challenges to data timeliness.

  \item \textbf{Velocity-Data Correctness:} It is vital to optimize the use of limited computing resources to transfer data [P7]. Data in the high-speed transmission process will greatly increase the data failure rate. Abnormal or missing data will affect the correctness and availability of big data applications [P54].
\end{itemize}

Overall, as summarized in Table~\ref{tab:test8}, the \emph{volume} property has a significant impact on all aspects of data quality, including \emph{data correctness}, \emph{data timeliness}, \emph{data completeness}, and so on. \emph{Variety} affects the \emph{data consistency}, \emph{data accuracy}, etc. \emph{Velocity} plays an important role in \emph{data timeliness} and \emph{data correctness}.

We mainly determine the influences of big data attributes on big data applications. Specifically, the major data quality issues are mostly because of \emph{volume}, and we have finalized five major quality parameters, including \emph{data timeliness}, \emph{data completeness}, \emph{data correctness}, \emph{data accuracy} and \emph{data consistency}.

\subsection{Identify the Important Quality Attributes in Big Data Applications (RQ2)}\label{sec:Identify the important quality attributes in big data applications Using SLR (RQ2)}

The goal of this section is to answer RQ2 ({\em which kind of important quality attributes do big data applications need to ensure?}). We use the ISO/IEC 25010:2011 to extract the  quality attributes of big data applications.

Table~\ref{tab:test9} provides the statistical distribution of different quality attributes. According to statistics, we identify related quality attributes, as shown in Figure~\ref{fig7}. For some articles that may involve more than one quality attribute, such as [P41] and [P58], we choose the main quality attribute that they convey.

From the 83 primary studies, some quality attributes are discussed, including \emph{correctness}, \emph{performance}, \emph{availability}, \emph{scalability}, \emph{reliability}, \emph{efficiency}, \emph{flexibility}, \emph{robustness}, \emph{stability}, \emph{interoperability}, and \emph{consistency}.

From Fig.~\ref{fig7}, we can see that the five main quality attributes have been discussed in the 83 primary studies. However, there are fewer articles focused on other attributes, such as \emph{stability}, \emph{consistency}, \emph{efficiency}, and so on. In addition, from the relevant literature, we can analyze which technologies can affect the corresponding quality attributes, although there is no clear statement in the relevant literature. Table~\ref{tab:test10-1} shows some of the technologies that may affect \emph{correctness}, \emph{performance}, \emph{availability}, \emph{scalability}, and \emph{reliability}. These technologies can help us to further understand these quality attributes. We will now focus on the main quality attributes.

\begin{table}[]
 \centering
\caption{\label{tab:test10-1}{Techniques for Addressing Quality Attributes}}
\begin{tabular}{p{1.5cm}p{6.6cm}}
 \toprule[1pt]
  \textbf{Attributes} & \textbf{Techniques}\\\midrule[1pt]
  Correctness & {Fault-tolerance mechanism} \\\hline
  Performance & {Parallel architecture, Multicloud cross-layer cloud monitoring framework, Cache, Model-driven architecture, Write buffer, Performance analysis model} \\\hline
  Availability & {Multicloud cross-layer cloud monitoring framework, Fault-tolerance mechanism, BigQueue} \\\hline
  Scalability & {Flexible data analytic framework, distributed storage system, bloat-aware design and so on} \\\hline
  Reliability & {Fault-tolerance mechanism, Heterogeneous NoSQL databases, Condition monitoring} \\\bottomrule[1pt]
\end{tabular}
\end{table}

\begin{itemize}
  \item \textbf{Correctness:} \emph{Correctness} measures the probability that big data applications can `get things right'. If the big data application cannot guarantee the correctness, then it will have no value at all. For example, a weather forecast system that always provides the wrong weather is obviously not of any use. Therefore, correctness is the first attribute to be considered in big data applications. The papers [P25] and [P23] provide the fault tolerance mechanism to guarantee the normal operation of applications. If the big data application runs incorrectly, it will cause inconvenience or even loss to the user. The papers [P40] and [P43] provide the testing method to check the fault of big data applications to assure the correctness.
  \item \textbf{Performance:} \emph{Performance} refers to the ability of big data applications to provide timely services, specifically in three areas, including the average response time, the number of transactions per unit time and the ability to maintain high-speed processing.
  However, \emph{volume}, \emph{variety} and \emph{velocity} attributes of big data attributes have an impact on these three aspects.
  Due to the large amounts of data, performance is a key topic in big data applications. In Table~\ref{tab:test9}, we show the many relative papers that refer to the performance of big data applications. The major purpose of focusing on the \emph{performance} problem is to handle big data with limited resources in big data applications. To be precise, the processing performance of big data applications under massive data scenarios is its major selling point and breakthrough. According to the relevant literature, we can see that common performance optimization technologies for big data applications are generally divided into two parts [P57], [P56], [P52], [P26], [P25], [P14]. The first one is hardware and system-level observations to find specific bottlenecks and make hardware or system-level adjustments. The second one is to achieve optimization mainly through adjustments to specific software usage methods.
  \item \textbf{Availability:} \emph{Availability} refers to the ability of big data applications to run without any issue for a long time. The rapid growth of data has made it necessary for big data applications to manage data streams and handle an impressive \emph{volume}, and since these data types are complex (\emph{variety}), the operation process may create different kinds of problems. Consequently, it is important to ensure the \emph{availability} of big data applications.
  \item \textbf{Scalability:} \emph{Scalability} refers to the ability of large data applications to maintain service quality when users and data volumes increase. The \emph{volume} of big data attributes will inevitably bring about the scalability issue of big data applications. Specifically, the scalability of big data applications includes system scalability and data scalability. For a continuous stream of big data, processing systems, storage systems, etc. should be able to handle these data in a scalable manner. Moreover, the system would be very complex for big data applications.
  For better improvement, the system must be scalable. Paper P7 proposes a flexible data analytic framework for big data applications, and the framework can flexibly handle big data with scalability.
  \item \textbf{Reliability:} \emph{Reliability} refers to the ability of big data applications to apply the specified functions within the specified conditions and within the specified time. \emph{Reliability} issues are usually caused by unexpected exceptions in the design and undetected code defects. For example, paper [P58] uses a monitoring technique to monitor the operational status of big data applications in real time so that failures can occur in real time and developers can effectively resolve these problems.
\end{itemize}

\begin{figure}[h]
  \centering
  \includegraphics[width=1\linewidth]{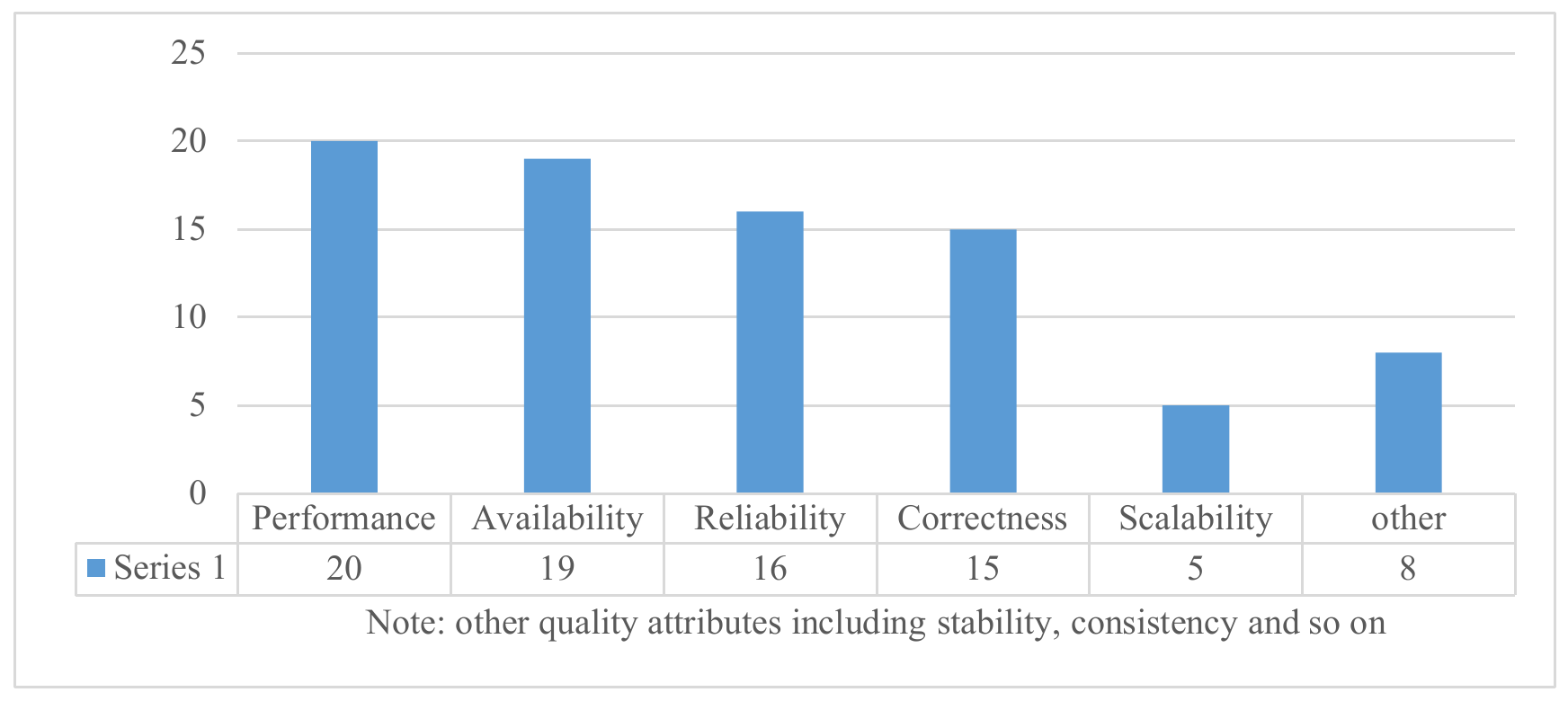}\\
  \caption{The Frequency Distribution of the Quality Attributes}\label{fig7}
\end{figure}

\begin{table}[]
 \centering
\caption{\label{tab:test9}Distribution of the Quality Attributes}
\begin{tabular}{p{1.5cm}p{6.2cm}}
  \toprule[1pt]
  \textbf{Attributes} & \textbf{Studies}\\\midrule[1pt]
  {Correctness} & {[P40], [P23], [P32], [P33], [P25], [P43], [P28], [P45], [P48], [P50], [P63], [P71], [P72], [P73], [P74]} \\\hline

  {Performance} & {[P39], [P41], [P55], [P26], [P57], [P66], [P34], [P35], [P8], [P1], [P36], [P4], [P64], [P11], [P19], [P62], [P70], [P75], [P77], [P78]} \\\hline

  {Availability} & {[P52], [P14], [P27], [P15], [P4], [P6], [P12], [P13], [P17], [P20], [P29], [P37], [P44], [P46], [P47], [P69], [P76], [P79], [P80]} \\\hline

  {Scalability} & {[P7], [P2], [P59], [P21], [P60]} \\\hline

  {Reliability} & {[P56], [P58], [P16], [P36], [P10], [P5], [P18], [P22], [P30], [P31], [P38], [P49], [P51], [P61], [P68], [P74]} \\\hline

  {Others} & {[P24], [P54], [P67], [P9], [P3], [P81], [P82], [P83]} \\

  \bottomrule[1pt]
\end{tabular}
\end{table}

Although big data applications have many other related quality attributes, the most important ones are the five mentioned above. Therefore, these five quality attributes are critical to ensuring the quality of big data applications and are the main focus of this survey.

\subsection{Technologies for Assuring the QA of Big Data Applications (RQ3)}\label{sec:Identify QA Techniques Using SLR (RQ3)}
This section answers RQ3 ({\em which kinds of technologies are used to guarantee the quality of big data applications?}). We extracted the quality assurance technologies used in the primary articles. In Fig.~\ref{fig8}, we show the distribution of papers for these different types of QA technologies. These technologies cover the entire development process for big data applications. According to the papers we collected, we identified eight technologies in our study, i.e.,  \emph{specification}, \emph{analysis}, \emph{model-driven architecture (MDA)}, \emph{fault tolerance}, \emph{testing}, \emph{verification}, \emph{monitoring}, and \emph{fault and failure prediction}.

\begin{figure}[ht]
\setcounter{figure}{7}
  \centering
  \includegraphics[width=1\linewidth]{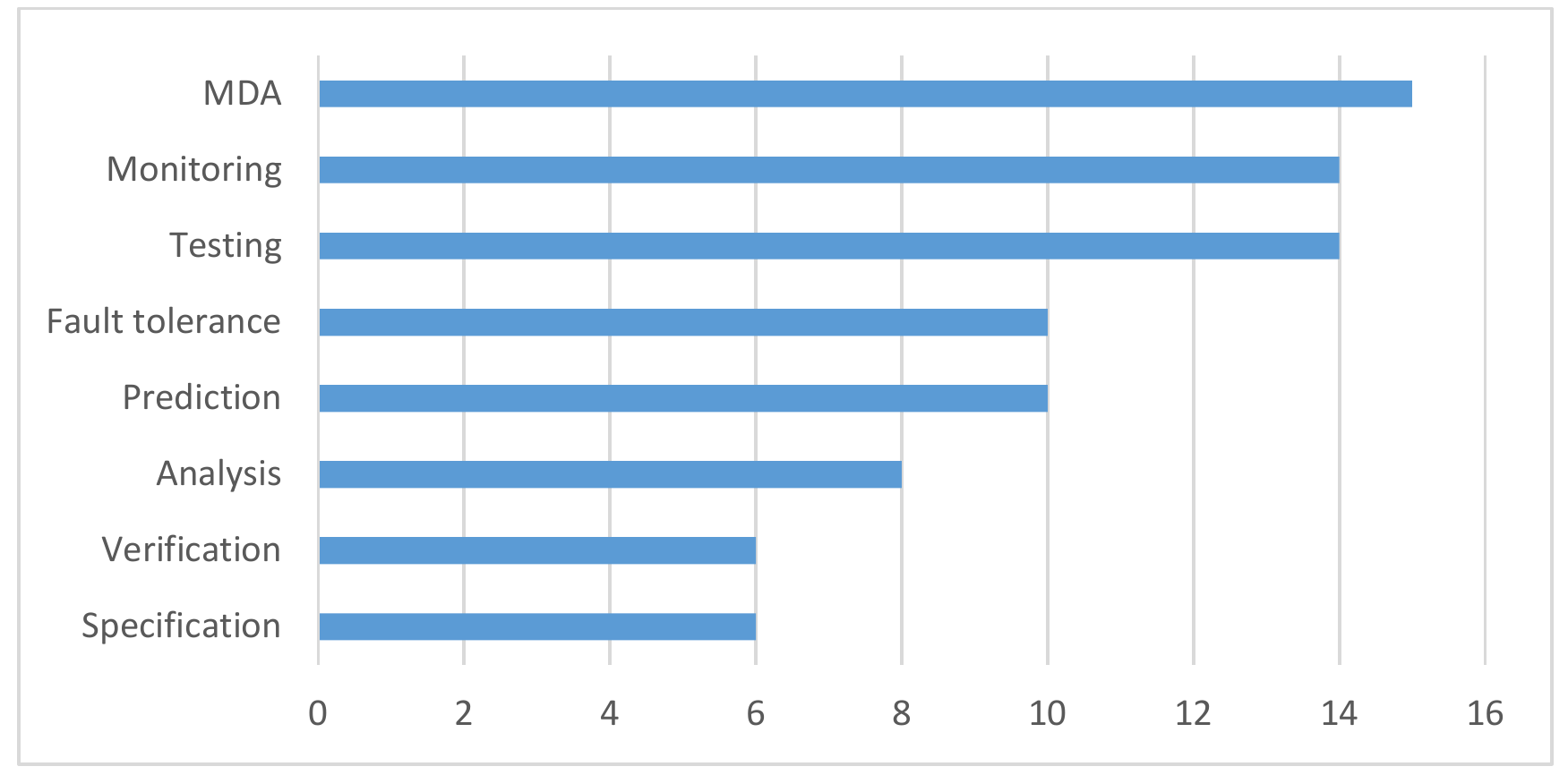}\\
  \caption{Distribution of Different QA Technologies}\label{fig8}
\end{figure}

In fact, quality assurance is an activity that applies to the entire big data application process.
The development of big data applications is a system engineering task that includes requirements, analysis, design, implementation, and testing.
Accordingly, we mainly divide the QA technologies into design time and runtime.
Based on the papers we surveyed and the development process of big data applications, we divide the quality assurance technologies for big data applications into the above eight types. In Table~\ref{tab:test10}, we have listed the simple descriptions of different types of QA technologies for big data applications. Detailed explanations of the eight technologies are provided in Appendix A.

\renewcommand\arraystretch{1.6}
\begin{table*}[!htb]
 \centering
\caption{\label{tab:test10}Identified QA Technologies}
\begin{tabular}{p{2.2cm}p{8cm}p{6cm}}
 \toprule[1pt]
  \textbf{QA technologies} & \textbf{Description} & \textbf{References}\\\toprule[1pt]
  {Specification} & {A specification refers to a type of technical standard to ensure the quality in the design time.} & {[P1], [P2], [P3], [P4], [P5], [P6]}\\\hline

  {Analysis} & {This technique can analyze the performance and other quality attributes of big data applications to determine the main factors that affect their quality.} & {[P7], [P8], [P9], [P10], [P11], [P12], [P13], [P75]}\\\hline

  {Model-Driven Architecture (MDA)} & {MDA provides a way (through related tools) to standardize a platform-independent application and select a specific implementation platform for the application, transforming application specifications to a specific implementation platform.} & {[P14], [P15], [P16], [P17], [P18], [P19], [P20], [P21], [P22], [P76], [P77], [P78], [P81], [P82], [P83]}\\\hline

  {Fault tolerance} & {Fault tolerance serves as an effective means to address reliability and availability concerns of big data applications.} & {[P23], [P24], [P25], [P26], [P27], [P28], [P29], [P30], [P31], [P32]}\\\hline

  {Verification} & {The purpose of verification is to verify that the design of the output ensures that the design phase of the input requirements is met.} & {[P33], [P34], [P35], [P36], [P37], [P38]}\\\hline

  {Testing} & {The purpose of testing is not only acknowledging application levels of correctness, performance and other quality attributes but also checking the testability of big data applications.} & {[P39], [P40], [P41], [P42], [P43], [P44], [P45], [P46], [P47], [P48], [P49], [P50], [P51], [P79]}\\\hline

  {Monitoring} & {Monitoring can detect failures or potential anomalies at runtime and is an effective method to guarantee the quality of big data applications.} & {[P52], [P53], [P54], [P55], [P56], [P57], [P58], [P59], [P60], [P61], [P62], [P63], [P64], [P80]}\\\hline

  {Fault and Failure Prediction} & {The goal is to achieve the prediction of the performance status and potential anomalies of big data applications and to provide the important and abundant information for real-time control.} & {[P65], [P66], [P67], [P68], [P69], [P70], [P71], [P72], [P73], [P74]}\\
  \bottomrule[1pt]
\end{tabular}
\end{table*}

\subsection{Existing Strengths and Limitations (RQ4)}\label{sec:Existing Strengths and Limitations (RQ4)}

The main purpose of RQ4 ({\em what are the advantages and limitations of the proposed technologies?}) is to comprehend the strengths and limitations of QA approaches. To answer RQ4, we first compare quality assurance technologies from six aspects. Finally, we discuss the strengths and limitations of each technique.

\subsubsection{Comparison}

We compare the quality assurance technologies from six different aspects, including \emph{suitable stage}, \emph{application domain}, \emph{quality attribute}, \emph{effectiveness}, \emph{usability} and \emph{efficiency}, according to the study published by Patel and Hierons~\cite{Patel2017A}.

\emph{Suitable stage} refers to the possible stage using these quality assurance technologies, including design time, runtime or both. \emph{Application domain} means the specific area that big data applications belong to, such as recommendations systems and prediction systems.
\emph{Quality attribute} identifies the most used quality attributes being addressed by the corresponding technologies. \emph{Effectiveness} means that big data application quality assurance technologies can guarantee the quality of big data applications to a certain extent. \emph{Usability} refers to the extent to which quality assurance technology can guarantee the quality of big data applications in the quality assurance environment of big data applications to achieve specific goals, such that effectiveness, efficiency, and satisfaction are achieved. \emph{Efficiency} refers to the improvement of the quality of big data application through the QA technologies. For better comparisons, we present the results in Table~\ref{tab:test11}\footnote{the QA are those presented in Table \ref{tab:test9}.}.

\renewcommand\arraystretch{1.5}
\begin{table*}[htbp]
 \centering
\caption{\label{tab:test11}Comparison of QA Technologies}
\begin{tabular}{p{1.6cm}p{1.3cm}p{2.2cm}p{1.6cm}p{3.8cm}p{2.5cm}p{2.6cm}}
 \toprule[1pt]
  \textbf{{Technologies}} & \textbf{{Stage}} & \textbf{{Application Domain}} & \textbf{{Quality Attributes}} & \textbf{Effectiveness} & \textbf{Usability} & \textbf{Efficiency}\\\midrule[1pt]
  {Specification } & {Design time } & {General} & {Efficiency, Performance, Scalability} & {Establishing detailed functional and behavioral descriptions, performance requirements, and so on to ensure the quality.} & {Ensure that all quality requirements of the user are met during the design phase.} & {Normally, specification can achieve highly efficient results.} \\\hline

  {Analysis} & {Design time } & {General} & {Performance, Scalability, Flexibility} & {Analyzes the main factors that affect the quality, provides a basis for testing of big data applications.} & {Analyze the quality impact factors of a given application as much as possible} & {According to the analysis results, different levels of efficiency can be achieved.} \\\hline

  {MDA} & {Design time } & {Hadoop Framework or MapReduce, Database application} & {Reliability, Efficiency, Performance, Availability} & {Use models to guide the design, development, and maintenance of systems, providing ease of big data application quality.} & {The model determines the quality of the follow-up.} & {It can achieve high efficiency due to the big data application framework.} \\\hline

  {Fault-tolerance} & {Design and Runtime} & {Hadoop, Distributed Storage System, Artificial Intelligence System} & {Performance, Correctness, Reliability, Scalability} & {Allow big data applications within a certain range to allow or tolerate the occurrence of mistakes. Even if a minor error occurs, the big data application can still offer stable operation. } & {Allows the system to make a small number of errors, effectively guaranteeing quality.} & {For any big data application with fault tolerance, its quality is bound to be better.} \\\hline

  {Verification} & {Design and Runtime} & {Big Data Classification System, MapReduce and so on} & {Correctness, Reliability, Performance} & {Verification is based on the test of the entire system analysis, especially functional analysis.} & {Identify features that should not be implemented and reduce the complexity of big data applications.} & {Depends on the verification approaches, different approaches achieve different efficiency effects.} \\\hline

  {Testing} & {Design and Runtime} & {Large Databases, Geographic Security Control\ System and Bioinformatics Software} & {Correctness, Scalability, Performance, Scalability} & {Under specified conditions, it can operate applications to detect errors, measure software quality, and evaluate whether they meet design requirements.} & {Detect the big data application error created in the development stage.} & {Depends on the testing method.} \\\hline

  {Monitoring} & {Runtime} & {Hadoop, MapReduce, and General} & {Performance, Availability, Reliability} & {Monitoring the occurrence of errors in big data applications and providing alerts that allow managers to spot errors and fix them.} & {Effectively monitor the occurrence of errors } & {Monitoring is only an approach that provides aid and can also lead to a high load. } \\\hline

  {Fault and Failure Prediction } & {Runtime} & {Cloud platforms, Automated IT System and so on} & {Reliability, Performance} & {Predicting errors that may occur in the operation of big data applications, so that errors can be prevented in advance.} & {May predict errors that were not discovered before.} & {Based on the prediction results, varying degrees of effectiveness can be achieved.} \\
  \bottomrule[1pt]
\end{tabular}
\end{table*}

From Table~\ref{tab:test11}, we can see that \emph{specification}, \emph{analysis}, \emph{verification} and \emph{MDA} are used at design time. \emph{Testing}, \emph{monitoring} and \emph{fault and failure prediction} are used at runtime. \emph{Fault tolerance} covers both design time and runtime. Design-time technologies are commonly used in MapReduce, which is a great help when designing big data application frameworks. The runtime technologies are usually used after the generation of big data applications, and their application is very extensive, including intelligent systems, storage systems, cloud computing, etc.

For quality attributes, while most technologies can contend with \emph{performance} and \emph{reliability}, some technologies focus on \emph{correctness}, \emph{scalability}, etc. To a certain extent, these eight quality assurance technologies assure the quality of big data applications during their application phase, although their \emph{effectiveness}, \emph{usability}, and \emph{efficiency} are different. To better illustrate these three quality parameters, we carry out a specific analysis through two quality assurance technologies. \emph{Specification} establishes complete descriptions of information, detailed functional and behavioral descriptions, and performance requirements for big data applications to ensure the quality. Therefore, it can guarantee the integrity of the function of big data applications to achieve the designated goal of big data applications and guarantee satisfaction. Although it does not guarantee the quality of the application runtime, it guarantees the quality of the application in the initial stage. As a vital technology in the QA of big data applications, the main function of testing is to test big data applications at runtime. As a well-known concept, the purpose of testing is to detect errors and measure quality, thus ensuring \emph{effectiveness} and \emph{usability}. In addition, the \emph{efficiency} of testing is largely concerned with testing methods and testing tools. The detailed description of other technologies is shown in Table~\ref{tab:test11}.

\subsubsection{Strengths and Limitations}

\begin{itemize}
  \item \textbf{Specification}: Due to the large amounts of data generated for big data applications, suitable specifications can be used to select the most useful data at hand. This technology can effectively improve the efficiency, performance and scalability of big data applications by using UML [P5], ADL [P4] and so on.
The quality of the system can be guaranteed at the design stage. In addition, the specification is also used to ensure that system functions in big data applications can be implemented correctly. However, all articles are aimed at a specific application or scenario, and do not generalize to different types of big data applications [P4], [P2].

  \item \textbf{Analysis}: The main factors that affect the big data applications' quality analysis are the size of the data, the speed of data processing, and data diversity. Analysis technologies can analyze major factors that may affect the quality of software operations during the design phase of big data applications. Current approaches only focus on analyzing performance attributes [P8], [P7]. There is a need to develop approaches for analyzing other quality attributes. In addition, it is impossible to analyze all quality impact factors for big data applications. The specific conditions should be specified before analysis.
  \item \textbf{MDA}: MDA uses a single model to generate and export most codes for big data applications and can greatly reduce human error. To date,  MDA metamodels and model mappings are only targeted for very special kinds of systems, e.g., MapReduce [P15], [P14].
Metamodels and model mapping approaches for other kinds of big data applications are also urgently needed.
  \item \textbf{Fault tolerance}: Fault tolerance is one of the staple metrics of quality of service in big data application. Fault-tolerant mechanisms permit big data applications to allow or tolerate the occurrence of mistakes within a certain range. If a minor error occurs, the big data application can still offer stable operation [P30], [P24]. Nevertheless, fault tolerance cannot always be optimal. Furthermore, fault tolerance can introduce performance issues, and most current approaches neglect this problem.
  \item \textbf{Verification}: Due to the complexity of big data applications, there is no uniform verification technology in general.
   Verification technologies verify or validate quality attributes by using logical analysis, theorem proving and model checking. There is a lack of formal models and corresponding algorithms to verify attributes of big data applications [P35], [P34]. Due to the existence of big data attributes, traditional software verification standards are no more meet the quality requirements\cite{Hussain2016Collect}.

  \item \textbf{Testing}: In contrast to verification, the testing technique is always performed during the execution of big data applications. Due to the large amounts of data, automatic testing is an efficient approach for big data applications.
  Current research always applies traditional testing approaches in big data applications [P45]. However, novel approaches for testing big data attributes are urgently needed because testing focuses are different between big data application testing and conventional testing. Conventional testing focuses on diverse software errors regarding structures, functions, UI, and connections to the external systems.
  In contrast, big data application testing focuses on involute algorithms, large-scale data input, complicated models and so on. Furthermore, conventional testing and big data application testing are different in the test input, the testing execution and the results.
  As an example, learning-based testing approaches [P73] can test the \emph{velocity} attribute of big data applications.
  \item \textbf{Monitoring}: Monitoring can obtain accurate status and behavior information for big data applications in a real operating environment. For big data applications running in a complex and variable network environment, their operating environment will affect the operation of the software system and produce some unexpected problems.
  Therefore, monitoring technologies will be more conducive to the timely response to the emergence of anomalies to prevent failures [P59], [P52]. A stable and reliable big data application relies on monitoring technologies that not only monitor whether the service is alive or not but also monitor the operation of the system and data quality. The high \emph{velocity} of big data engendered the challenge of monitoring accuracy issues and may produce overhead problems for big data applications.
  \item \textbf{Fault and Failure Prediction}: Prediction technologies can predict errors that may occur in the operation of big data applications so that errors can be prevented in advance. Due to the complexity of big data applications, the accuracy of prediction is still a substantial problem that we need to consider in the future. Deep learning-based approaches [P28], [P26], [P25] can be combined with other technologies to improve prediction accuracy due to the large amounts of data.
\end{itemize}

\subsection{Empirical Evidence (RQ5)}\label{sec:Empirical Evidence (RQ5)}

The goal of RQ5 ({\em what are the real cases of using the proposed technologies?}) is to elicit empirical evidence on the use of QA technologies. We organize the discussion along the QA technologies discussed in Section \ref{sec:Identify QA Techniques Using SLR (RQ3)}.
\renewcommand\arraystretch{1.4}
\begin{table*}[p]
 \centering
\caption{\label{tab:testnew}Experimental summary and statistics}
\begin{tabular}{p{0.6cm}p{2cm}p{1.9cm}p{3.6cm}p{3.8cm}p{4cm}}
 \toprule[1pt]
  \textbf{\tabincell{c}{Ref}} & \textbf{\tabincell{c}{Techniques}} & \textbf{\tabincell{c}{Case Type}} & \textbf{\tabincell{c}{Domain}} & \textbf{Metric} & \textbf{Experimental Results}\\\toprule[1pt]
  \tabincell{c}{[P4]} & \tabincell{c}{Specification} & \tabincell{c}{Small Case} & \tabincell{c}{Traffic forecasting system} & \tabincell{c}{Delay time} & \tabincell{c}{Rigorous, easy and expressive} \\\hline
  \tabincell{c}{[P7]} & \tabincell{c}{Analysis} & \tabincell{l}{Real-world \\Case} & \tabincell{l}{Scientific data compression \\and remote visualization} & \tabincell{c}{Latency, Throughput} & \tabincell{c}{Obtain two factors} \\\hline
  \tabincell{c}{[P8]} & \tabincell{c}{Analysis} & \tabincell{c}{Large Case} & \tabincell{c}{MapReduce application} & \tabincell{l}{Processing time, Job \\turnaround, Hard disk \\bytes written} & \tabincell{c}{Improve the performance} \\\hline
  \tabincell{c}{[P15]} & \tabincell{c}{MDA} & \tabincell{c}{Small  Case} & \tabincell{l}{Analytics-intensive big \\data applications} & \tabincell{l}{Accidental complexities, \\Cycle} & \tabincell{l}{The effectiveness of MDD \\for accidental complexities} \\\hline
  \tabincell{c}{[P18]} & \tabincell{c}{MDA} & \tabincell{c}{Small Case} & \tabincell{c}{Not mentioned} & \tabincell{l}{CPU, memory, network \\utilization levels} & \tabincell{c}{Improve the scalability} \\\hline
  \tabincell{c}{[P19]} & \tabincell{c}{MDA} & \tabincell{c}{Small Case} & \tabincell{c}{Word Count application} & \tabincell{l}{Not mentioned} & \tabincell{c}{High degree of automation} \\\hline
  \tabincell{c}{[P26]} & \tabincell{c}{Fault tolerance} & \tabincell{l}{Real-world \\Case} & \tabincell{l}{Join bidirectional data \\streams} & \tabincell{l}{Time Consuming, \\Recover Ratio} & \tabincell{l}{Improve the performance of \\joining two-way streams} \\\hline
  \tabincell{c}{[P28]} & \tabincell{c}{Fault tolerance} & \tabincell{l}{Real-world \\Case} & \tabincell{l}{MapReduce data \\computing applications} & \tabincell{l}{CPU utilization, Memory \\footprint, Disk throughput, \\Network throughput} & \tabincell{l}{Transparently enable fault\\tolerance for applications} \\\hline
  \tabincell{c}{[P41]} & \tabincell{c}{Testing} & \tabincell{c}{Small Case} & \tabincell{l}{Network public opinion \\monitoring system} & \tabincell{c}{Response time} & \tabincell{c}{No specific instructions} \\\hline
 \tabincell{c}{[P50]} & \tabincell{c}{Testing} & \tabincell{c}{Small Case} & \tabincell{c}{Image Processing} & \tabincell{c}{Error detection rate } & \tabincell{l}{ Detects all the embedded \\mutants} \\\hline
  \tabincell{c}{[P33]} & \tabincell{c}{Verification} & \tabincell{c}{Small Case} & \tabincell{c}{Cell Morphology Assay} & \tabincell{c}{MRs} & \tabincell{l}{Its effectiveness for testing \\ADDA} \\\hline
  \tabincell{c}{[P56]} & \tabincell{c}{Monitoring} & \tabincell{l}{Real-world \\Case} & \tabincell{c}{Cloud monitoring system} & \tabincell{c}{Insert operation time} & \tabincell{l}{Achieves a response time of a \\few hundred} \\\hline
  \tabincell{c}{[P55]} & \tabincell{c}{Monitoring} & \tabincell{l}{Real-world \\Case} & \tabincell{l}{Big data public opinion \\monitoring platform} & \tabincell{l}{Accuracy Rate, Elapsed \\time /MS} & \tabincell{l}{High accuracy and meeting \\the requirements of real time} \\\hline
  \tabincell{c}{[P53]} & \tabincell{c}{Monitoring} & \tabincell{c}{Large Case} & \tabincell{c}{No specific instructions} & \tabincell{l}{Throughput, Read latency, \\Write latency} & \tabincell{l}{Improve the performance \\by changing various tuning \\parameters} \\\hline
  \tabincell{c}{[P58]} & \tabincell{c}{Monitoring} & \tabincell{c}{Small Case} & \tabincell{l}{Big data-based \\condition monitoring of \\power apparatuses} & \tabincell{c}{No specific instructions} & \tabincell{l}{Improve the accuracy of \\condition monitoring} \\\hline
  \tabincell{c}{[P67]} & \tabincell{c}{Prediction} & \tabincell{c}{Small Case} & \tabincell{c}{Cloud computing system} & \tabincell{l}{False Positive Rate, true \\positive rate} & \tabincell{l}{Achieve high true positive \\rate, low false positive \\rate for failure prediction} \\\hline
  \tabincell{c}{[P65]} & \tabincell{c}{Prediction} & \tabincell{c}{Small Case} & \tabincell{c}{Different applications} & \tabincell{c}{Prediction accuracy} & \tabincell{l}{New prediction system is \\accurate and efficient} \\\bottomrule[1pt]
\end{tabular}
\end{table*}

\begin{itemize}
  \item \emph{Specification.} The approach in [P4] is explained by a case study of specifying and modeling a Vehicular Ad hoc NETwork (VANET). The major merits of the posed method are its capacity to take into consideration big data attributes and cyber physical system attributes through customized concepts and models in a strict, simple and expressive approach.
  \item \emph{Analysis.} The experiments in [P7] show that the two factors that are the most important concern the quality of scientific data compression and remote visualization, which are analyzed by latency and throughput.
Experiments in [P8] were conducted to analyze the connection between the performance measures of several MapReduce applications and performance concepts, such as CPU processing time. The consequences of performance analysis illustrate that the major performance measures are processing time, job turnaround and so on. Therefore, in order to improve the performance of big data applications, we must take into consideration these measures.
  \item \emph{MDA.} In [P15], the investigators demonstrate the effectiveness of the proposed approach by using a case study. This approach can overcome accidental complexities in analytics-intensive big data applications. Paper P18 conducts a series of tests using Amazon's AWS cloud platform to evaluate performance and scalability of the observable architecture by considering the CPU, memory, and network utilization levels. Paper P19 uses a simple case study to evaluate the proposed architecture and a metamodel in the Word Count Application.
  \item \emph{Fault tolerance.} The experiments in paper [P26] show that DAP architecture can improve the performance of joining two-way streams by analyzing the time consumption and recover ratio. In addition, all data can be reinstated if the newly started VMs can be reinstated in a few seconds while nonadjacent nodes fail; meanwhile, if neighboring nodes fail, some data can be reinstated. Through analyzing the CPU utilization, memory footprint, disk throughput and network throughput, experiments in paper [P28] show that the performance of all cases (MapReduce data computing applications) can be significantly improved.
  \item \emph{Verification.} In [P33], the author uses CMA(cell morphology assay) as an example to describe the design of the framework. Verifying and validating datasets, software systems and algorithms in CMA demonstrates the effectiveness of the framework.
  \item \emph{Testing.} In [P41], the authors use a number of virtual users to simulate real users and observe the average response time and CPU performance in a network public opinion monitoring system. In [P50], the experiment verifies the effectiveness and correctness of the proposed technique in alleviating the Oracle problem in a region growth program. The testing method successfully detects all the embedded mutants.
  \item \emph{Monitoring}. The experiments in [P56] show that a large queue can increase the write speed and that the proposed framework supports a very high throughput in a reasonable amount of time in a cloud monitoring system. The authors also provide comparative tests to show the effectiveness of the framework. In [P55], the comparison experiment shows that the method is reliable and fast, especially with the increase of the data volume, and the speed advantage is obvious.
  \item \emph{Fault and Failure Prediction.} In [P67], the authors implement the proactive failure management system and test the performance in a production cloud computing environment. Experimental results show that the approach can reach a high true positive rate and a low false positive rate for failure prediction. In [P65], the authors provide emulation-based evaluations for different sets of data traces, and the results show that the new prediction system is accurate and efficient.
\end{itemize}

\section{Discussion}\label{sec:DISCUSSION}
The key findings are already provided in Table~\ref{tab:test1}. In this Section, we mainly discuss the cross-cutting findings and existing challenges of this review.

\subsection{Cross-cutting Findings}\label{sec:Cross-cutting Findings}
This subsection discusses some cross-cutting findings deduced from the key findings.
\begin{itemize}
\item \emph{Relations between big data attributes and quality attributes}. The collected results based on main findings show that big data attributes sometimes have contradictory impacts on quality attributes. Some big data attributes are found to improve some quality attributes and weaken others. These findings lead to a conclusion that big data attributes do not always improve all quality attributes of big data applications. To a certain extent, this conclusion matches the definition of big data attributes stated by most of the researchers involved in a study regarding the challenges and benefits between big data attributes and quality attributes in practice.
\item \emph{Relations among big data attributes, quality attributes and big data applications}. In this study, researchers have proposed some quality attributes to effectively assess the impact of big data attributes on applications. Therefore, we believe that it is incorrect to limit the research on the quality of big data applications to a certain big data property, obtain some negative results, and then state a general conclusion that comprehensive consideration of big data attributes causes big data applications' quality to weaken. For example, considering that the data that the system need to process has a large \emph{volume}, fast \emph{velocity}, and huge \emph{variety}, a number of companies have built sophisticated monitoring and analyzing tools that go far beyond simple resource utilization reports. The continuous improvement of monitoring and analyzing tools enables big data application systems to occupy more resources, which means longer \emph{response times} and lower \emph{performance}[P53]. Consequently, most big data applications that take into account big data attributes can cause the system to be more complex. We believe that it is incorrect to draw a general conclusion that comprehensive consideration of big data attributes negatively affects big data applications' quality. Moreover, such a conclusion does not consider other quality attributes such as \emph{reliability}, \emph{scalability}, and \emph{correctness}.
\item \emph{Relations between quality attributes and QA technologies}. It is important to note that researchers may also use different QA technologies when considering the same quality attributes. That is, there may be some empirical experience enlisted in practice. It can be inferred that the relations between quality attributes and QA technologies is not one-to-one. For example, \emph{correctness} can be achieved through a variety of QA technologies, including \emph{testing}, \emph{fault tolerance}, \emph{verification}, \emph{monitoring}, and \emph{fault and failure prediction}, as analyzed from Tables~\ref{tab:test9} and~\ref{tab:test10}. On the other hand, when using the same QA technology, different researchers design different methods and evaluation indicators. Therefore, when a study finds that there is a negative or a positive relation between quality attributes and QA technologies, we cannot conclude a specific finding regarding the relation between them.
    We need to consider investigating this problem for various types of big data applications.
    For example, for big data applications concerned with privacy, monitoring technologies can improve the reliability of the system, but in some common big data applications, an excessive emphasis on monitoring technologies may degrade the performance of the system. Therefore, the specification of QA technologies and the relationship between QA technologies and quality attributes need to be further studied.
\end{itemize}
\subsection{Existing Challenges}\label{sec:Existing Challenges (RQ5)}
Based on the key findings and cross-cutting findings aforementioned, we discuss some research challenges in this subsection.

\begin{itemize}
  \item \textbf{Challenge 1: The existing problems brought by big data attributes.}
\end{itemize}

Despite that many technologies have been proposed to address big data attributes, existing technologies cannot provide adequate scalability and face major difficulties.
Based on the SLR results, Table~\ref{tab:test12} summarizes the challenges and the possible solutions for 3V attributes. For example, the distributed file system has a high fault tolerance, high throughput and other excellent characteristics. It can use multiple storage servers to share the storage load to store a large amount of data and support linear expansion. When the storage space is insufficient, it can use hot swap to increase storage devices and expand conveniently. These capabilities address the storage and scalability challenges of big data applications caused by the \emph{volume} attribute. Many studies~\cite{Elkafrawy2017HDFSX, Radha2014Efficient} show that distributed file systems can handle large-scale data very well.

For large-scale optimization and high-speed data transmission of big data applications, a decomposition-based distributed parallel programming algorithm~\cite{Ke2016On} is proposed and an online algorithm is designed to dynamically adjust data partitioning and aggregation. Dobre et al.~\cite{Dobre2014Parallel} review various parallel and distributed programming paradigms, analyze how they fit into the big data era, and present modern emerging paradigms and frameworks. Consequently, parallel programming is particularly effective in big data applications, especially for addressing the\emph{velocity} of data.
In addition, the NoSQL~\cite{Reniers2017On} database is created to solve the challenges brought by the multiple data types of large-scale data collection, especially the big data application problems. NoSQL's flexible storage structure fundamentally solves the problem of \emph{variety} and unstructured data storage [P23]. At the same time, \emph{distributed file systems} solve the problem of data storage and greatly reduce costs. It can be seen that these technologies can be combined with existing QA technologies for big data applications in the future.

\renewcommand\arraystretch{1.7}
\begin{table}[]
 \centering
\caption{\label{tab:test12}Properties, Challenges and Technologies}
\begin{tabular}{lll}
  \toprule[1pt]
  \textbf{Properties} & \textbf{Challenge} & \textbf{Possible Solutions} \\\midrule[1pt]
  Volume & \tabincell{c}{Storage/Scale} & \tabincell{c}{Distributed File Systems} \\\hline
  Velocity & \tabincell{c}{Fast Processing} & \tabincell{c}{Parallel Programming} \\\hline
  Variety & \tabincell{c}{Heterogeneity} & \tabincell{c}{NoSQL Databases} \\
  \bottomrule[1pt]
\end{tabular}
\end{table}

\begin{itemize}
  \item \textbf{Challenge 2: Lack of the awareness and good understanding of QA technologies for big data applications.}
\end{itemize}

As mentioned in Section~\ref{sec:Cross-cutting Findings}, because different professional skills and understanding of the field exist, big data practitioners tend to use different QA technologies when considering the same quality attributes; therefore, the QA technologies that are chosen according to experience may not be the most appropriate. Moreover, an incorrect application of QA technologies may cause extensive losses. For example, because of an incorrect transaction algorithm, the electronic trading system led to the purchase of 150 different stocks at a low price by the United States KCP (Knight Capital Group) financial companies, resulting in the company suffering a loss of 440 million US dollars with the day shares falling 62\%\footnote{https://dealbook.nytimes.com/2012/08/02/knight-capital-says-trading-mishap-cost-it-440-million/}. Therefore, a clear understanding of QA technologies can reduce the implementation of incorrect algorithms and technologies in big data applications, thus avoiding huge losses. Nevertheless, the variety and diversity of big data applications makes it difficult to enact a theory of QA technologies to normalize them, which creates the challenge regarding a lack of awareness of QA technologies. In general, fully understanding the capabilities and limitations of QA technologies can address the specific needs of big data applications. Consequently, researchers are advised to fill this gap by deeply exploring theoretical research, considering more mature QA technologies, and making use of the studies frequently applied in practice.

\begin{itemize}
  \item \textbf{Challenge 3: Lack of quantitative models and algorithms to measure the relations among big data attributes, data quality parameters and software quality attributes.}
\end{itemize}

\begin{figure}
  \centering
  \includegraphics[width=1\linewidth]{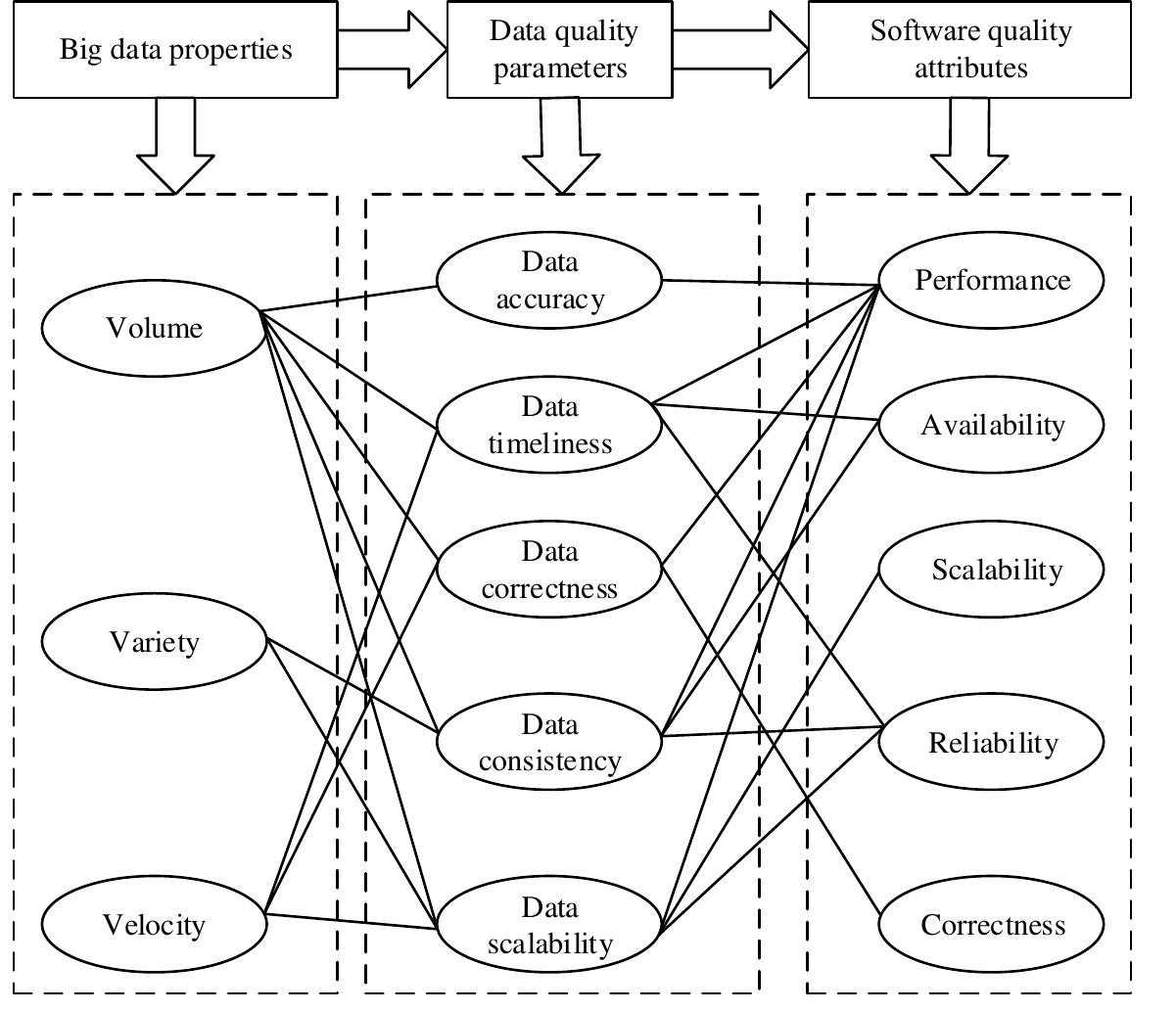}\\
  \caption{Relations among Big Data Properties, Data Quality Parameters and Software Quality Attributes}\label{fig9}
\end{figure}

The SLR results show that big data attributes are related to the quality of software. However, big data attributes should first affect multiple data quality parameters; then, the quality of data attributes affects the quality of software. Figure~\ref{fig9} shows our primary study on the relations among big data attributes, data quality parameters, and software quality attributes. However, the change of an attribute is often accompanied by the change of multiple attributes. More detailed theories, models and algorithms are needed to precisely understand the different kinds of relations. To specify quality requirements in the context of big data applications, paper P1 presents a novel approach to address some unique requirements of engineering challenges in big data to specify quality requirements in the context of big data applications. The approach intersects big data attributes with software quality attributes, and then it identifies the system quality requirements that apply to the intersection. Nevertheless, the approach is still in the early stages and has not been applied to the development environment of big data applications. Hence, it is still a considerable challenge and a trending research issue.

\renewcommand\arraystretch{1.2}
\begin{table}[ht]
 \centering
\caption{\label{tab:test13}3V Properties, Software Quality Attributes and Techniques of Big Data Application}
\begin{tabular}{p{2.6cm}p{2.4cm}p{2.5cm}}
  \toprule[1pt]
  \textbf{3V properties} & \textbf{\tabincell{l}{Software quality \\attributes}} & \textbf{Technologies} \\\midrule[1pt]
  Velocity, Variety, Volume & \tabincell{c}{Reliability} & \tabincell{c}{Specification} \\\hline
  Velocity, Volume & \tabincell{c}{Performance} & \tabincell{c}{Analysis} \\\hline
  Volume & \tabincell{l}{Performance, \\Scalability} & \tabincell{c}{MDA} \\\hline
  Variety, Volume & \tabincell{l}{Performance, \\Scalability} & \tabincell{c}{Fault tolerance} \\\hline
  Volume, Variety & \tabincell{l}{Performance, \\Reliability} & \tabincell{c}{Verification} \\\hline
  Variety, Velocity & \tabincell{l}{Availability, \\Performance} & \tabincell{c}{Testing} \\\hline
  Variety, Velocity & \tabincell{l}{Performance, \\real time} & \tabincell{c}{Monitoring} \\\hline
  Variety & \tabincell{l}{Performance, \\Dependability} & \tabincell{l}{Fault and\\ Failure Prediction} \\
  \bottomrule[1pt]
\end{tabular}
\end{table}

\begin{itemize}
  \item \textbf{Challenge 4: Lack of mature tools for QA technologies for big data applications.}
\end{itemize}

In Section~\ref{sec:Identify QA Techniques Using SLR (RQ3)}, we have summed up eight QA technologies for big data applications based on the selected 83 primary studies. Nevertheless, many authors discussed existing limitations and needed improvements. Therefore, existing technologies can solve quality problems to a certain extent. From Table~\ref{tab:test13}, we can see that the 3V properties will result in software quality issues, and the corresponding technologies can partially address those problems.

However, with the wide application of machine learning in the field of big data application, the quality attributes of big data application gradually appear some new non functional attributes, such as fairness and interpretability. Processing a large amount of data needs to consume more resources of the application system. The performance of big data application is an urgent matter to be considered. Fair distribution of the resources of big data application can greatly improve the quality of big data application. For example, in distributed cloud computing, storage and bandwidth resources are usually limited, and such resources are usually very expensive, so collaborative users need to use resources fairly [P81]. In [P82], an elastic online scheduling framework is proposed to guarantee big data applications fairness. Another attribute is interpretability. Interpretability is a subjective concept which is hard to reach a consensus. Considering the two different dimensions of semantics and complexity, when dealing with big data, researchers often pay attention to the performance indicators of big data applications, such as accuracy, but these indicators can only explain some problems, and the black box part of big data applications is still difficult to explain clearly. Due to the rapid increase of the amount of data to be processed as time goes by, the structure of big data application will gradually become more complex, which increases the difficulty of interpretation system. At this time, it is very important to study the interpretability of big data applications[P83]. However, we have collected few papers about these non functional attributes so far, and the research is still in its infancy, lacking mature tools or technologies.

In addition, although there are many mature QA tools for traditional software systems, none of the surveyed approaches discusses any mature tools that are dedicated to big data applications. Indeed, if practitioners want to apply QA technologies for big data applications today, they would have to implement their own tool, as there are no publicly available and maintained tools. This is also a very significant obstacle for the widespread use of QA technologies for big data applications in empirical research as well as in practice.

\section{Threats to Validity}\label{sec:THREATS TO VALIDITY}

In the design of this study, several threats are encountered. Similar to all SLR studies, a common threat to validity regards the coverage of all relevant studies. In the following, we discuss the main threats of our study and the ways we mitigated them.

\textbf{External validity:} In the data collection process, most of the data are collected by three researchers; this may lead to incomplete data collection, as some related articles may be missing. Although all authors have reduced the threat by determining unclear questions and discussing them together, this threat still exists. In addition, each researcher may be biased and inflexible when he extracts the data, so at each stage of the study, we have ensured that at least two other reviewers have reviewed the work. Another potential threat is the consideration of studies which are only published in English. However, since the English language is the main used language for academical papers, this threat is considered as as minimal reasonably.

\textbf{Internal validity:} This SLR may have missed some related novel research papers. To alleviate this threat, we have searched for papers in big data-related journals/conferences/workshops. In total, 83 primary studies are selected by using the SLR. The possible threat is that QA technologies are not clearly shown for the selected primary studies. In addition, we endeavored as much as possible to extract information to analyze each article, which helps to avoid missing important information. This approach can minimize the threats as much as possible.

\textbf{Construct validity:} This concept relates to the validity of obtained data and the research questions. For the systematic literature review, it mainly addresses the selection of the main studies and how they represent the population of the questions. We have taken steps to reduce this threat in several ways. For example, the automatic search is performed on several electronic databases so as to avoid the potential biases.

\textbf{Reliability:} Reliability focuses on ensuring that the results are the same if our review would be conducted again. Different researchers who participated in the survey may be biased in collecting and analyzing data. To solve this threat, the two researchers simultaneously extracted and analyzed data strictly according to the screening strategy, and further discussed the differences of opinions in order to enhance the objectivity of the research results. Nevertheless, the background and experience of the researchers may have produced some prejudices and introduced a certain degree of subjectivity in some cases. This threat is also related to replicating the same findings, which in turn affects the validity of the conclusion.

\section{Conclusions and Future Work}\label{sec:Conclusions and Future Work}
In this paper, we conducted a systematic literature review on QA technologies for big data applications. It mainly discusses the state-of-art technologies to ensure the quality of big data applications. Based on this, we present five research questions. To reach our goal, we applied a database search approach to identify the most relevant studies on the topic of study, and 83 primary studies are selected. Finally, we analyze the data collected from these studies to answer the research questions presented earlier in Section~\ref{sec:Research Questions}.

Using the SLR, a list of eight QA technologies has been identified. These technologies not only play an important role in the research of big data applications but also impact actual big data applications. Although researchers have proposed these technologies to ensure quality, the research on big data quality is, however, still in its infancy stage, and problems regarding quality still exist in big data applications.
The results of this study are useful for future research in QA technologies for big data applications.
Based on our discussions, the following topics may be part of our future work:
\begin{itemize}
  \item Considering quality attributes with big data properties together to ensure the quality of big data applications.
  \item Understanding and tapping into the limitations, advantages and applicable scenarios of QA technologies.
  \item Researching quantitative models and algorithms to measure the relations among big data properties, data quality attributes and software quality attributes.
  \item Developing mature tools to support QA technologies for big data applications.
\end{itemize}

\bibliographystyle{elsarticle-num}
\bibliography{refs1}

\clearpage
\begin{appendix}
    \section{}
    \begin{itemize}
    \item \textbf{Specification}\label{sec:Specification}
    A specification refers to a type of technical standard. The specification includes requirements specification, functional specification, and non-functional specification of big data applications. Paper P3 thinks that many useless data may lead to a lot of time for data processing. Only the correct specifications can collect valuable data and improve the efficiency of big data applications. Paper P4 proposes an approach which integrates Architecture Analysis \& Design Language (AADL) to consider big data properties through customized concepts and models in a rigorous way. Paper P5 uses UML 2.3 with formal methods to standardize cloud-based applications. The semantic models can fully describe cloud security functions and ensure the secrecy of data which are stored in the cloud. Paper P6 defines a vocabulary that can integrate security requirements specifications into cloud architecture based on the unified modeling language model, which significantly improves the productivity and success rate of distributed cloud application and system development. Paper P1 presents a novel approach to address requirement engineering challenges in big data scopes. The approach intersects the big data properties with quality attributes, and then it identifies the system quality requirements that apply to the intersection. Paper P2 proposes a bloat-ware design paradigm for the development of efficient and scalable big data applications in object-oriented Garbage Collection enabled languages.
\end{itemize}

\begin{itemize}
    \item \textbf{Analysis}\label{sec:Analysis}

    The analysis technique is used to analyze the main factors which can affect the big data application quality such as performance, correctness, and others, which plays an important role in the design phase of big data applications.

    Paper P13 conducts a comprehensive survey on threat analysis techniques of existing software systems and analyze their applicability. Paper P8 proposes a performance analysis model based on the concept of metrology and aspects of software quality that is directly related to performance. Paper P7 proposes a framework to find the best strategy to reduce data movement in specific situations created by the large volume of data. Paper P12 restores the architecture of the system and guarantees the quality of the data intensive systems through different methods based on static analysis. Paper P9 proposes the data confidentiality challenges in big data applications. Paper P10 uses the fault tree model to model the reliability of big data applications on the cloud. The fault tree analysis (FTA) technology can provide good references for the fault processing as well as quality assurance of big data applications. Paper P11 focuses on security aspects of data acquisition, transmission, storage, decision making and other control phases. The approach is validated on some industrial big data applications with security risk. Paper P75 develops a system called OntoDMA, which is based on the cloud environment via big data analysis technique.
\end{itemize}

\begin{itemize}
    \item \textbf{Model-Driven Architecture (MDA)}\label{sec:MDA}

    Model-driven architecture (MDA) refers to quality assurance (QA) techniques for helping developers creating software that meets not functional properties, such as, reliability, safety and efficiency [P16].

    Paper P14 proposes an extension tool which can automate the integration of the Hadoop platform. The main idea is to divide each problem into several sub-tasks based on a simple programming model (MapReduce). Paper P21 proposes an approach via MDA to support semi-automatic development of big data applications. Due to the fact that existing models and QA techniques often ignore big data properties osuch as volume and velocity, paper P16 presents the research agenda of DICE (DICE aims at defining a framework for the quality-driven development of Big Data applications). To overcome accidental complexities of big data application, paper P15 presents a MapReduce meta-model and proposes a model-driven development approach for MapReduce based applications which can improve the availability and reliability of a system. Paper P22 designs, validates, and develops MapReduce applications on cloud systems using MDA which can ensure the robustness and reliability of services from early design to runtime. Paper P20 describes a model-driven and step-by-step approach for efficient and reliable design or optimization based on data intensive applications executed over clouds. Paper P18 presents an architecture that automates metric collection processes, and a model-driven approach is used to configure a distributed framework at runtime. Paper P19 designs an architecture to support MDA for big data applications, explaining with a case study.

    Paper P76 provides a goal-oriented architecture to reduce the risk of wrong decision.
    Paper P77 proposes a management framework to continuously monitor and manage the operation of big data system, which ensures the stability of the whole system.
    Paper P78 systematically investigates big data tools, technologies( such as Hadoop and Spark), case studies on distributed machine learning tools such as Mahout, Spark MLlib and so on.
    Paper P81 proposes a novel optimization framework for fair allocation of resources in a distributed cloud environment, which is illustrated by conducting experiments by simulations.
    Paper P82 proposes an elastic and online scheduling framework for big data applications.
    Paper P83 optimizes the rule base as well as the fuzzy set parameters, a distributed multi-objective evolutionary learning scheme that has also been extended.
\end{itemize}

\begin{itemize}
    \item \textbf{Fault-tolerance}\label{sec:Fault-tolerance}

    Paper P26 presents a novel architecture called dual-assembly-pipeline (DAP) which enabled with fault-tolerance technique. To avoid losing data in case of hardware failure, paper P25 suggests introducing a cache in the distributed storage system. Paper P32 proposes a new fault-tolerant mechanism supporting iterative graph processing for distributed big data systems, which reduces checkpoint cost and recovery time. Paper P27 discusses typical fault features in cloud services and fault-tolerant solutions that are useful for system reliability and availability. Paper P23 poses a data migration method that is efficient and fault-tolerance, and it can ensure that the result is correct.
    Paper P24 introduces the fault-tolerance in big data including batch computing, spark, and SDN due to the big data properties. Paper P29 presents BeTL (Beneath the Task Level) which adds slight changes to MapReduce, and makes it possible to support fault-tolerance. 
    Paper P28 proposes a Library-level fault tolerance mechanism with global persistent state model, which can transparently implement application fault tolerance in the framework of MPI and MapReduce. Paper P30 proposes a novel data-driven integrated solution, which combines learning and optimizing ability of the monitored data to assure the fault-tolerance and also ensure the efficiency of the systems.
\end{itemize}

\begin{itemize}
    \item \textbf{Verification}\label{sec:Verification}

    Verification relies on existing technologies to validate whether the big data applications satisfy desired quality attributes. Many challenges for big data applications appear due to the 3V properties. For example, the \emph{volume} and the \emph{velocity} of data may bring a difficult task to validate the correctness attribute of big data applications.

    To solve the challenges brought by \emph{volume} and \emph{velocity} of big data, paper P33 has developed a big data system called CMA (Cell Morphology Assay), and have developed a framework that rigorously can verify large-scale image data software systems and the corresponding machine learning algorithms. Paper P34 presents MtMR (merkle tree-based MapReduce), which utilize existing verification method to verify the integrity of MapReduce applications. Their analysis shows that MtMR can give a high result integrity while incurring a moderate performance overhead. Paper P35 proposes a dynamic security framework via a shared key which is driven from some synchronized numbers. Paper P37 proposes a distributed approach which exploits cloud computing facilities to verify complex systems. Paper P38 has worked with BDA application developers in industry for three years to find that traditional verification methods cannot solve the runtime and debugging problems of the applications during deployment phase. Consequently, they propose a lightweight approach which can uncover the differences between some pseudo and the large-scale cloud deployments. In this way, the efficiency and accuracy of deployment verification are improved.
\end{itemize}

\begin{itemize}
    \item \textbf{Testing}\label{sec:Testing}

    The SLR results show that different kinds of testing techniques are used to validate whether big data applications conform to requirements specification. Using this techniques, inconsistent or contradictory with the requirements specification can be identified.

    Paper P42 gives an overview of the technologies and challenges of existing system testing. By analyzing the changes of data attributes, it comes to the conclusion that because of pure data and various kinds of data, it is impossible to test manually and only can test big data application automatically. To address these problems, Paper P40 proposes a novel big data framework using testing technology, which can improve the correctness and scalability of the applications.

    The performance plays a particularly important role in big data applications. Performance testing is a typical test technique for non-functional testing. Paper P41 proposes a novel performance testing technology, which provides testing goal, analysis, design, and load design for big data applications. Paper P39 presents the challenges and techniques for testing. They describe the need of testing in big data, methodologies for testing of big data, problems faced and challenges in the testing of big data in detail. Paper P43 poses that the search-based software testing techniques that can be extended to big data applications. Paper P45 presents two black-box testing techniques that automatically detect faults. In addition, they performed an empirical study to evaluate the effectiveness and efficiency of the testing techniques proposed (MRTest Random and MRTest-t-Wise) compared to the MRUnit. The method does not need to know the expected output during the test process, which is very suitable for testing of big data applications.
    According to an evolving set of features and usage patterns, constantly updated data often leads to ``outdated" testing. Paper P46 presents an automated approach to validate performance tests by comparing workload signatures from tests and the field using execution logs. Paper P51 describes the current situation of the software testing technology under the background of big data based on the UML model. Paper P47 proposes a software system test method based on a formal model. The use of formal models for software system testing has the advantages of a clear test process, easy determination of test adequacy standards, suitability for static detection, and ease of automation. To solve the high input and output scale, paper P48 proposes an AI software testing method based on scene deduction method. Due to the uncertainty of operation and large volume of data, it is difficult to obtain the test oracle of image region growth program. For mobile bank Apps testing in Hybrid-Cloud, paper P49 presents a cloud-based testing framework, which can improve the Apps quality. To alleviate the test oracle problem, paper P50 applies the metamorphic testing method into image region growth program testing and proposes a series of metamorphic relations by analyzing the geometric properties.
    Paper P79 proposes a new system in order to test three key properties, availability, consistency and efficiency (ACE) of SQL and NoSQL systems.

\end{itemize}

\begin{itemize}
    \item \textbf{Monitoring}\label{sec:Monitoring}

    Plenty of structured, unstructured, and semi-structured data are generated in big data applications. These data are very complex, huge, as well as rapidly changing. If the data is not filtered, it is difficult to achieve runtime monitoring. As a result, one of the biggest challenges for such applications is the real-time analysis and processing of the data. Runtime monitoring is an effective way to ensure the overall quality of big data applications. However, Runtime monitoring may occur additional loading problems for big data applications. Hence, it is necessary to improve the performance of big data monitoring.

    To improve the overall performance of monitoring technology, paper P55 presents a dual cloud architecture that can take the full advantage of network bandwidth and cloud resources. Paper P59 proposes an adaptive algorithm for monitoring big data applications to accommodate updated intervals, data characteristics and administrator requirements. Paper P63 proposes a general purpose MAD (monitoring, alerting and diagnosis) system for big data applications in order to keep data freshness and reduce maintenance cost. Paper P57 summarizes the existing monitoring techniques and tools in big data. In addition, they particularly focus on performance monitoring. Paper P54 introduces the data quality issues and discuss ways to monitor and control data quality. Because of the flood of data generated from smart phones and sensors, paper P52 proposes a service framework for the implementation of cloud surveillance. This framework supports real-time QoS monitoring to ensure the availability and performance of big data applications. Paper P64 proposes a system-level monitoring method for a heterogeneous big data streaming system to flexibly meet the quality of the systems used in different environments.
    Paper P56 proposes a monitoring framework based on big data solutions for efficiently storing and managing the data. This makes monitoring big data applications a challenging task due to the lack of standards and techniques for modeling and analysis of execution data (i.e., logs) generated by big data applications. Paper P61 presents their monitoring solution that performs real-time fault detection in big data applications. They analyze the complexity, scalability and availability of the proposed solution, which proves that it can efficiently monitor in big data applications. Based on the characteristics of big data and big data of power system, paper P58 proposes an integrated condition monitoring system of the transformer, GIS and power cable to assess system status. The technique greatly improves the accuracy of condition monitoring and diagnostics of power equipment.
    Paper P80 uses the Elasticsearch system to monitor and analyze big data application.
\end{itemize}

\begin{itemize}
    \item \textbf{Fault and Failure Prediction}\label{sec:Prediction}

    Big data applications faces many failures. If the upcoming failure can be predicted, he overall quality of big data applications may be greatly improved.

    To address this problem, Paper P72 makes a detailed failure analysis on unsuccessful execution and develops three online prediction models to improve the accuracy of fault prediction for big data systems. Paper P74 proposes a fast data update protocol for distributed storage systems that handle big data. The fault prediction mechanism proposed in this protocol improves the data update performance by 30\%\ and the prediction accuracy reaches 80\%\. Paper P66 proposes a method to disk failure prediction that combines manufacturer telemetry data with performance counters of the server and then converts them and use a two-stage machine learning model to achieve a high degree of accuracy in predicting disk failures. Paper P67 presents an unsupervised failure detection approach using Bayesian models. It can not only characterize the normal system execution states but also can detect the anomalous behaviors. Paper P68 designs a prediction system based on a recurrent neural network (LSTM), which is a log-driven failure prediction system.
    Many machine learning methods apply to fault prediction in big data application, paper P69 aims at the comparison of 14 machine learning techniques for development of effective defect prediction models. Paper P73 analyses and compares six kinds of machine learning fault prediction methods, and confirmed the predictive ability of machine learning methods for software fault prediction in big data applications. To solve the macro integrity of software programs in software defect prediction and the correlation between local defects and surrounding program elements, paper P71 introduces complex network technology into defect prediction, establishes a software network model, and uses complex network metrics to design a set of defects that can reflect local and global measure of features, and a dynamic predictive model based on threshold filtering algorithm is proposed.
\end{itemize}

\renewcommand\arraystretch{1.4}
\begin{table*}
\setcounter{table}{16}
\section{}
 \centering
 \setlength{\belowcaptionskip}{0.2cm}
 \caption{\label{tab:test16}Selected papers}
\begin{supertabular}{p{0.5cm}p{4cm}p{0.5cm}p{6cm}p{5cm}}
 \toprule[1pt]
  \textbf{Paper} & \textbf{Authors} & \textbf{Year} & \textbf{Title} & \textbf{Source}\\\toprule[1pt]
  {P1} & {Ibtehal Noorwali, Darlan Arruda} & {2016} & {Understanding quality requirements in the context of big data systems} & {International Workshop on BIG Data Software Engineering} \\\hline

  {P2} & {Yingyi Bu, Vinayak Borkar, Guoqing Xu, Michael J. Carey.} & {2013} & {A bloat-aware design for big data applications} & {ACM SIGPLAN Notices} \\\hline

  {P3} & {Noufa Al-Najran, Ajantha Dahanayake} & {2015} & {A Requirements Specification Framework for Big Data Collection and Capture} & {East European Conference on Advances in Databases and Information Systems} \\\hline

  {P4} & {Lichen Zhang} & {2014} & {A framework to specify big data driven complex cyber physical control systems} & {IEEE International Conference on Information and Automation} \\\hline

  {P5} & {Kunding Fang, Xiaohong Li, Jianye Hao, Zhiyong Feng} & {2016} & {Formal modeling and verification of security protocols on cloud computing systems based on UML 2.3} & {IEEE International Conference On Big Data Science And Engineering} \\\hline

  {P6} & {Massimo Ficco, Francesco Palmieri, Aniello Castiglione} & {2015} & {Modeling security requirements for cloud-based system development} & {Concurrency and Computation Practice and Experience} \\\hline

  {P7} & {Hongbo Zou, Yongen Yu, Wei Tang, Hsuan-Wei Michelle Chen} & {2014} & {FlexAnalytics: A Flexible Data Analytics Framework for Big Data Applications with I/O Performance Improvement} & {Big Data Research} \\\hline

  {P8} & {Luis Eduardo Bautista Villalpando, Alain April} & {2014} & {Performance analysis model for big data applications in cloud computing} & {Journal of Cloud Computing} \\\hline

  {P9} & {Jian Yin, Dongfang Zhao} & {2015} & {Data confidentiality challenges in big data applications} & {IEEE International Conference on Big Data} \\\hline

  {P10} & {Rui Cao, Jing Gao} & {2018} & {Research on reliability evaluation of big data system} & {IEEE International Conference on Cloud Computing and Big Data Analysis} \\\hline

  {P11} & {Xiaorong Feng, Shizhun Jia, Songtao Mai} & {2018} & {The research on industrial big data information security risks} & {IEEE International Conference on Big Data Analysis} \\\hline

  {P12} & {Csaba Nagy} & {2013} & {Static Analysis of Data-Intensive Applications} & {European Conference on Software Maintenance and Reengineering} \\\hline

  {P13} & {Katja Tuma, Gul Calikli, Riccardo Scandariato} & {2018} & {Threat Analysis of Software Systems: A Systematic Literature Review} & {Journal of Systems and Software} \\\hline

  {P14} & {Mohammed Alodib, Zaki Malik} & {2015} & {A Big Data approach to enhance the integration of Access Control Policies for Web services} & {IEEE International Conference on Computer and Information Science} \\\hline

  {P15} & {Asha Rajbhoj, Vinay Kulkarni, Nikhil Bellarykar} & {2014} & {Early Experience with Model-Driven Development of MapReduce Based Big Data Application} & {Software Engineering Conference} \\\hline

  {P16} & {Giuliano Casale, Danilo Ardagna, Matej Artac} & {2015} & {DICE: Quality-Driven Development of Data-Intensive Cloud Applications} & {IEEE International Workshop on Modeling in Software Engineering}  \\\hline

  {P17} & {Noriko Etani} & {2015} & {Database application model and its service for drug discovery in Model-driven architecture} & {Journal of Big Data}\\\bottomrule[1pt]

\end{supertabular}
\end{table*}

\clearpage
\renewcommand\arraystretch{1.3}
\begin{table*}[]
 \centering
\begin{supertabular}{p{0.5cm}p{4cm}p{0.5cm}p{6cm}p{5cm}}
 \toprule[1pt]
  \textbf{Paper} & \textbf{Authors} & \textbf{Year} & \textbf{Title} & \textbf{Source}\\\toprule[1pt]

  {P18} & {John Klein, Ian Gorton, Laila Alhmoud, Joel Gao, Caglayan} & {2016} & {Model-Driven Observability for Big Data Storage} & {IEEE Conference on Software Architecture} \\\hline

  {P19} & {Michele Guerriero, Saeed Tajfar, Damian Andrew Tamburri} & {2016} & {Towards a model-driven design tool for big data architectures} & {International Workshop on Big Data Software Engineering} \\\hline

  {P20} & {Rafael Tolosana-Calasanz, Jose Angel Banares, Jose-Manuel Colom} & {2018} & {Model-driven development of data intensive applications over cloud resources} & {Future Generation Computer Systems} \\\hline

  {P21} & {S. Sousa Osvaldo Jr, Denivaldo Lopes, Aristófanes C. Silva, Zair Abdelouahab} & {2017} & {Developing Software Systems to Big Data Platform based on MapReduce model: an Approach based on Model Driven Engineering} & {Information and Software Technology} \\\hline

  {P22} & {Flora Amatoa, Francesco Moscatob} & {2016} & {Model transformations of MapReduce Design Patterns for automatic development and verification} & {Journal of Parallel and Distributed Computing} \\\hline

  {P23} & {Marco Scavuzzo, Damian A, Tamburri, Elisabetta Di Nitto} & {2016} & {Providing Big Data Applications with Fault-Tolerant Data Migration across Heterogeneous NoSQL Databases} & {International Workshop on BIG Data Software Engineering} \\\hline

  {P24} & {Xing Wu, Zhikang Du, Shuji Dai, Yazhou Liu} & {2017} & {The Fault Tolerance of Big Data Systems} & {Communications in Computer and Information Science} \\\hline

  {P25} & {Lars Lundberg, Hakan Grahn, Dragos Ilie, Christian Melander} & {2015} & {Cache Support in a High Performance Fault-Tolerant Distributed Storage System for Cloud and Big Data} & {IEEE International Parallel and Distributed Processing Symposium Workshops} \\\hline

  {P26} & {Xinchun Liu, Xiaopeng Fan, Jing Li} & {2013} & {A Novel Parallel Architecture with Fault-Tolerance for Joining Bi-Directional Data Streams in Cloud} & {International Conference on Cloud Computing and Big Data} \\\hline

  {P27} & {R. Jhawar, V. Piuri} & {2014} & {Fault Tolerance and Resilience in Cloud Computing Environments} & {Computer and Information Security Handbook} \\\hline

  {P28} & {Jian Lin, Fan Liang, Xiaoyi Lu, Li Zha, Zhiwei Xu} & {2015} & {Modeling and Designing Fault-Tolerance Mechanisms for MPI-Based MapReduce Data Computing Framework} & {IEEE First International Conference on Big Data Computing Service and Applications} \\\hline

  {P29} & {Hao Wang, Haopeng Chen, Zhenwei Du, Fei Hu} & {2017} & {BeTL: MapReduce Checkpoint Tactics Beneath the Task Level} & {IEEE Transactions on Services Computing} \\\hline

  {P30} & {Amin Majd, Elena Troubitsyna} & {2017} & {Data-Driven Approach to Ensuring Fault Tolerance and Efficiency of Swarm Systems} & {IEEE International Conference on Big Data} \\\hline

  {P31} & {Hongliang Li, JieWu, Zhen Jiang, Xiang Li, Xiaohui Wei} & {2017} & {Minimum Backups for Stream Processing With Recovery Latency Guarantees} & {IEEE Transactions on Reliability} \\\hline

  {P32} & {Chen Xu, Markus Holzemer, Manohar Kauly, Volker Markl} & {2016} & {Efficient fault-tolerance for iterative graph processing on distributed dataflow systems} & {IEEE International Conference on Data Engineering} \\\hline

  {P33} & {Junhua Ding, Xinhua Hu, Venkat Gudivada} & {2017} & {A Machine Learning Based Framework for Verification and Validation of Massive Scale Image Data} & {IEEE Transactions on Big Data} \\\hline

  {P34} & {Yongzhi Wang, Yulong Shen, Hua Wang, Jinli Cao} & {2016} & {MtMR: Ensuring MapReduce Computation Integrity with Merkle Tree-based Verifications} & {IEEE Transactions on Big Data} \\\bottomrule[1pt]

  \end{supertabular}
\end{table*}

\clearpage
\renewcommand\arraystretch{1.5}
\begin{table*}[]
 \centering
\begin{supertabular}{p{0.5cm}p{4cm}p{0.5cm}p{6cm}p{5cm}}
 \toprule[1pt]
  \textbf{Paper} & \textbf{Authors} & \textbf{Year} & \textbf{Title} & \textbf{Source}\\\toprule[1pt]

  {P35} & {Deepak Puthal, Surya Nepal, Rajiv Ranjan, Jinjun Chen} & {2016} & {DLSeF: A Dynamic Key-Length-Based Efficient Real-Time Security Verification Model for Big Data Stream} & {ACM Transactions on Embedded Computing Systems} \\\hline

  {P36} & {Chang Liu, Chi Yang, Xuyun Zhang, Jinjun Chen} & {2015} & {External integrity verification for outsourced big data in cloud and IoT} & {Future Generation Computer Systems} \\\hline

  {P37} & {Matteo Camilli} & {2014} & {Formal verification problems in a big data world: towards a mighty synergy} & {Companion International Conference on Software Engineering} \\\hline

  {P38} & {Weiyi Shang, Zhenming Jiang, Hadi Hemmati, Bram Adams} & {2013} & {Assisting developers of Big Data Analytics Applications when deploying on Hadoop clouds} & {International Conference on Software Engineering} \\\hline

  {P39} & {Naveen Garg, Sanjay Singla, Surender Jangra} & {2016} & {Challenges and Techniques for Testing of Big Data} & {Procedia Computer Science} \\\hline

  {P40} & {Nan Li, Anthony Escalona, Yun Guo, Jeff Offutt} & {2015} & {A Scalable Big Data Test Framework} & {IEEE International Conference on Software Testing} \\\hline

  {P41} & {Zhenyu Liu} & {2014} & {Research of performance test technology for big data applications} & {IEEE International Conference on Information and Automation} \\\hline

  {P42} & {Harry M. Sneed, Katalin Erdoes} & {2015} & {Testing big data (Assuring the quality of large databases)} & {IEEE International Conference on Software Testing} \\\hline

  {P43} & {Erik M. Fredericks, Reihaneh H. Hariri} & {2017} & {Extending search-based software testing techniques to big data applications} & {IEEE International Workshop on Search-Based Software Testing} \\\hline

  {P44} & {Raghunath Nambiar, Meikel Poess, Akon Dey, Paul Cao} & {2014} & {Introducing TPCx-HS: The First Industry Standard for Benchmarking Big Data Systems} & {Technology Conference on Performance Evaluation and Benchmarking} \\\hline

  {P45} & {Jesus Moran, Antonia Bertolino, Claudio de la Riva, Javier Tuya} & {2018} & {Automatic Testing of Design Faults in MapReduce Applications} & {IEEE Transactions on Reliability} \\\hline

  {P46} & {Mark D. Syer, Weiyi Shang, Zhenming Jiang} & {2017} & {Continuous validation of performance test workloads} & {Automated Software Engineering} \\\hline

  {P47} & {Weixiang Zhang, Wenhong Liu, Bo Wei} & {2017} & {Software system testing method based on formal model} & {IEEE  International Conference on Cloud Computing and Big Data Analysis} \\\hline

  {P48} & {Xinghan Zhao, Xiangfei Gao} & {2018} & {An AI Software Test Method Based on Scene Deductive Approach} & {IEEE Conference on Software Quality, Reliability and Security Companion} \\\hline

  {P49} & {Chaonian Guo, Shaomin Zhu, Tongsen Wang, Hao Wang} & {2018} & {FeT: Hybrid Cloud-Based Mobile Bank Application Testing} & {IEEE Conference on Software Quality, Reliability and Security Companion} \\\hline

  {P50} & {Chao Jiang, Song Huang, Zhanwei Hui} & {2018} & {Metamorphic Testing of Image Region Growth Programs in Image Processing Applications} & {IEEE Conference on Software Quality, Reliability and Security Companion} \\\bottomrule[1pt]

  {P51} & {Jing Wang, Dayong Ren} & {2018} & {Research on Software Testing Technology Under the Background of Big Data} & {IEEE Advanced Information Management, Communicates, Electronic and Automation Control Conference} \\\bottomrule[1pt]

  \end{supertabular}
\end{table*}

\clearpage
\renewcommand\arraystretch{1.3}
\begin{table*}[]
 \centering
\begin{supertabular}{p{0.5cm}p{4cm}<{\raggedright}p{0.5cm}p{6cm}p{5cm}}
 \toprule[1pt]
  \textbf{Paper} & \textbf{Authors} & \textbf{Year} & \textbf{Title} & \textbf{Source}\\\toprule[1pt]

  {P52} & {Khalid Alhamazani, Rajiv Ranjan, Prem Prakash Jayaraman} & {2014} & {Real-Time QoS Monitoring for Cloud-Based Big Data Analytics Applications in Mobile Environments} & {IEEE  International Conference on Mobile Data Management} \\\hline

  {P53} & {Tilmann Rabl, Mohammad Sadoghi, HansArno Jacobsen} & {2012} & {Solving big data challenges for enterprise application performance management} & {Proceedings of the VLDB Endowment} \\\hline

  {P54} & {Benjamin T.Hazen, ChristopherA.Boone, JeremyD.Ezell, L.AllisonJones-Farmer} & {2014} & {Data quality for data science, predictive analytics, and big data in supply chain management: An introduction to the problem and suggestions for research and applications} & {International Journal of Production Economics} \\\hline

  {P55} & {Hongwei Wang, Guowei Shi} & {2016} & {Research on Big Data Real-Time Public Opinion Monitoring under the Double Cloud Architecture} & {IEEE Second International Conference on Multimedia Big Data} \\\hline

  {P56} & {Saeed Zareian, Marios Fokaefs, Hamzeh, Xi Zhang} & {2016} & {A big data framework for cloud monitoring} & {International Workshop on BIG Data Software Engineering} \\\hline

  {P57} & {Gabriel Iuhasz, Ioan Dragan} & {2015} & {An Overview of Monitoring Tools for Big Data and Cloud Applications} & {International Symposium on Symbolic and Numeric Algorithms for Scientific Computing} \\\hline

  {P58} & {Jinxin Huang, Lin Niu, Jie Zhan} & {2015} & {Technical aspects and case study of big data based condition monitoring of power apparatuses} & {Power and Energy Engineering Conference} \\\hline

  {P59} & {Mauro Andreolini, Michele Colajanni, Marcello Pietri} & {2015} & {Adaptive, scalable and reliable monitoring of big data on clouds} & {Journal of Parallel and Distributed Computing} \\\hline

  {P60} & {Jonatan Enes, Roberto R. Exposito, Juan Tourino} & {2018} & {BDWatchdog: Real-time monitoring and profiling of Big Data applications and frameworks} & {Future Generation Computer Systems} \\\hline

  {P61} & {M. Omair Shafiq, Maryam Fekri, Rami Ibrahim} & {2017} & {MapReduce based Classification for Fault Detection in Big Data Applications} & {IEEE International Conference on Machine Learning and Applications} \\\hline

  {P62} & {Mauro Andreolini, Marcello Pietri, Stefania Tosi} & {2014} & {Monitoring Large Cloud-Based Systems} & {International Conference on Cloud Computing and Services Science} \\\hline

  {P63} & {Mingruo Shi, Ruiping Yuan} & {2015} & {MAD: A Monitor System for Big Data Applications} & {Intelligence Science and Big Data Engineering.} \\\hline

  {P64} & {Holger Eichelberger} & {2018} & {Flexible System-Level Monitoring of Heterogeneous Big Data Streaming Systems} & {Euromicro Conference on Software Engineering and Advanced Applications} \\\hline

  {P65} & {Dong Dai, Yong Chen, Dries Kimpe, Rob Ross} & {2015} & {Provenance-based object storage prediction scheme for scientific big data applications} & {IEEE International Conference on Big Data} \\\hline

  {P66} & {Sandipan Ganguly, Ashish Consul} & {2016} & {A Practical Approach to Hard Disk Failure Prediction in Cloud Platforms: Big Data Model for Failure Management in Datacenters} & {IEEE Second International Conference on Big Data Computing Service and Applications} \\\hline

  {P67} & {Qiang Guan, Ziming Zhang, Song Fu} & {2012} & {Ensemble of Bayesian Predictors and Decision Trees for Proactive Failure Management in Cloud Computing Systems} & {Journal of Communications} \\\hline

  {P68} & {Ke Zhang, Jianwu Xu, Martin Renqiang Min} & {2017} & {Automated IT system failure prediction: A deep learning approach} & {IEEE International Conference on Big Data} \\\bottomrule[1pt]

\end{supertabular}
\end{table*}

\clearpage
\renewcommand\arraystretch{1.6}
\begin{table*}[]
 \centering
\begin{supertabular}{p{0.5cm}p{4cm}<{\raggedright}p{0.5cm}p{6cm}p{5cm}}
 \toprule[1pt]
  \textbf{Paper} & \textbf{Authors} & \textbf{Year} & \textbf{Title} & \textbf{Source}\\\toprule[1pt]

  {P69} & {Ruchika Malhotra, Laavanye Bahl} & {2017} & {Empirical comparison of machine learning algorithms for bug prediction in open source software} & {International Conference on Big Data Analytics and Computational Intelligence} \\\hline

  {P70} & {Chao Shen, Weiqin Tong, Kim-Kwang Raymond Choo} & {2017} & {Performance prediction of parallel computing models to analyze cloud-based big data applications} & {Cluster Computing} \\\hline

  {P71} & {Yiwen Yang, Jun Ai, Fei Wang} & {2018} & {Defect Prediction Based on the Characteristics of Multilayer Structure of Software Network} & {IEEE Conference on Software Quality, Reliability and Security Companion} \\\hline

  {P72} & {Andrea Rosa, Walter Binder} & {2017} & {Failure Analysis and Prediction for Big-Data Systems} & {IEEE Transactions on Services Computing} \\\hline

  {P73} & {Ruchika Malhotra} & {2014} & {Comparative analysis of statistical and machine learning methods for predicting faulty modules} & {Applied Soft Computing Journal} \\\hline

  {P74} & {G. J. Akash, Ojus Thomas Lee, Sd Madhu Kumar} & {2017} & {RAPID: A Fast Data Update Protocol in Erasure Coded Storage Systems for Big Data} & {IEEE/ACM International Symposium on Cluster} \\\hline

  {P75} & {Hsuehyuan Lin, Shengyuan Yang} & {2019} & {A cloud-based energy data mining information agent system based on big data analysis technology} & {Microelectronics Reliability} \\\hline

  {P76} & {Mahdi Fahmideh, Ghassan Beydoun} & {2019} & {Big data analytics architecture design¡ªan application in manufacturing systems} & {Computers and Industrial Engineering} \\\hline

  {P77} & {Kwok Leung Tsui, Yang Zhao, Dong Wang} & {2019} & {Big Data Opportunities: System Health Monitoring and Management} & {IEEE Access} \\\hline

  {P78} & {T.Ramalingeswara Rao, Pabitra Mitra, Ravindara Bhatt, A.Goswami} & {2019} & {The big data system, components, tools, and technologies: a survey} & {Knowledge and Information Systems} \\\hline

  {P79} & {Maria Teresa Gonzalez Aparicio, Muhammad Younas} & {2019} & {Evaluation of ACE properties of traditional SQL and NoSQL big data systems} & {ACM/SIGAPP Symposium on Applied Computing} \\\hline

  {P80} & {Vlad Andrei Zamfir, Mihai Carabas, Costin Carabas, Nicolae Tapus} & {2019} & {Systems Monitoring and Big Data Analysis Using the Elasticsearch System} & {IEEE International Conference on Control Systems and Computer Science} \\\hline

  {P81} & {Qiufen Xia, Weifa Liang, Albert Y. Zomaya} & {2019} & {Collaboration- and Fairness-Aware Big Data Management in Distributed Clouds} & {IEEE Transactions on Information Forensics and Security} \\\hline

  {P82} & {Dawei Sun, Hongbin Yan, Shang Gao, Xunyun Liu, Rajkumar Buyya} & {2019} & {Rethinking elastic online scheduling of big data streaming applications over high-velocity continuous data streams} & {Journal of Supercomputing} \\\hline

  {P83} & {Marco Barsacchi, Alessio Bechini, Pietro Ducange, Francesco Marcelloni} & {2019} & {Optimizing Partition Granularity, Membership Function Parameters, and Rule Bases of Fuzzy Classifiers for Big Data by a Multi-objective Evolutionary Approach} & {Cognitive Computation} \\\bottomrule[1pt]

\end{supertabular}
\end{table*}

\end{appendix}

\end{document}